\documentclass[aps,prx,superscriptaddress,twocolumn,showpacs, longbibliography]{revtex4-1}
\usepackage{amsmath, amsthm, amssymb}
\usepackage{graphicx}
\usepackage{color}
\usepackage{times}

\usepackage{natbib}
\usepackage{hyperref}
\hypersetup{
        colorlinks=true,
}

\newcommand{\eqnref}[1]{(\ref{#1})}

\newcommand{\di}{\mathrm{d}}

\newcommand{\ua}{\uparrow}
\newcommand{\da}{\downarrow}
\renewcommand{\vec}[1]{{\mathbf #1}}

\newcommand{\comments}[1]{}
\newcommand{\concomp}[2]{\big\langle \! \big\langle{#1},{#2}\big\rangle \! \big\rangle}
\DeclareMathOperator{\sgn}{sgn}

\begin{document}

\title{Interplay between Kondo and Majorana interactions in quantum dots}

\author{Meng Cheng}
\affiliation{Station Q, Microsoft Research, Santa Barbara, CA 93106-6105, USA}
\affiliation{Condensed Matter Theory Center, Department of Physics, University of Maryland, College Park, MD 20742, USA}

\author{Michael Becker}
\affiliation{Institute for Theoretical Physics, University of Cologne, 50937 Cologne, Germany}

\author{Bela Bauer}
\affiliation{Station Q, Microsoft Research, Santa Barbara, CA 93106-6105, USA}

\author{Roman M. Lutchyn}
\affiliation{Station Q, Microsoft Research, Santa Barbara, CA 93106-6105, USA}

\begin{abstract}
We study the properties of a quantum dot coupled to a topological superconductor and a normal lead and discuss the interplay between Kondo and Majorana-induced couplings in quantum dot. The latter appears due to the presence of Majorana zero-energy modes localized, for example, at the ends of the one-dimensional superconductor. We investigate the phase diagram of the system as a function of Kondo and Majorana interactions using a renormalization-group analysis, a slave-boson mean-field theory and numerical simulations using the density-matrix renormalization group method. We show that, in addition to the well-known Kondo fixed point, the system may flow to a new fixed point controlled by the Majorana-induced coupling which is characterized by non-trivial correlations between a localized spin on the dot and the fermion parity of the topological superconductor and normal lead. We compute several measurable quantities such as differential tunneling conductance and impurity spin susceptibility which highlight some peculiar features characteristic to the Majorana fixed point.
\end{abstract}


\date{\today}
\maketitle

\section{Introduction}
Topological superconductors have recently attracted enormous theoretical and experimental interest~\cite{Reich, Brouwer_Science, Wilczek2012} because they can host certain exotic defects (e.g. vortices) that bind Majorana zero-energy modes. This excitement stems from the fact that such defects obey non-Abelian braiding statistics~\cite{Moore1991, Nayak1996, ReadGreen, Ivanov}, and can be utilized for topological quantum computing~\cite{TQCreview}. Among the many proposed realizations of topological superconductivity~\cite{Fu:2008, MajoranaQSHedge, Sau, Alicea, 1DwiresLutchyn, 1DwiresOreg, CookPRB'11, SauNature'12, YazdaniPRB'13}, a particularly promising one involves a quasi-one-dimensional semiconductor covered by an s-wave superconductor~\cite{1DwiresLutchyn, 1DwiresOreg}. Such a system effectively realizes the so-called Majorana quantum wire~\cite{Kitaev:2001} with Majorana zero-energy modes appearing at the opposite ends of the wire. First experimental signatures for Majorana zero-energy modes in semiconductor/superconductor heterostructure were shown by Mourik {\it et al.}~\cite{Mourik2012} using tunneling transport measurements. The appearance of a zero bias conduction peak characteristic for Majorana zero-energy modes (Majoranas) was observed at a finite magnetic field in agreement with theoretical predictions~\cite{ZeroBiasAnomaly0,ZeroBiasAnomaly1,ZeroBiasAnomaly2,ZeroBiasAnomaly3,ZeroBiasAnomaly31, ZeroBiasAnomaly4,ZeroBiasAnomaly5,ZeroBiasAnomaly6, 1DwiresLutchyn2, ZeroBiasAnomaly61, ZeroBiasAnomaly7}. This observation has excited the physics community since Majoranas can be manipulated in network structures of quasi-1D wires~\cite{AliceaBraiding, SauWireNetwork, ClarkeBraiding, TopologicalQuantumBus, BraidingWithoutTransport}, which opens up the possibility for topological quantum computing~\cite{kitaev, Freedman98, TQCreview}.

Inspired by this recent experimental progress~\cite{Mourik2012, Rokhinson2012, Das2012, Deng2012, Fink2012, Churchill2013, Chang_PRL2013, Lee_arxiv2013, Deng_arxiv2014}, we consider here a topological superconductor (TSC)-quantum dot (QD)-normal lead (NL) junction. Such structures might naturally form in the semiconductor nanowire experiments~\cite{Mourik2012, Das2012, Fink2012, Churchill2013} or can be purposely engineered using other potential experimental realizations of Majoranas ({\it e.g.}, the domain walls on the edge of 2D topological insulator~\cite{MajoranaQSHedge}) in order to control and manipulate Majoranas~\cite{Flensberg_PRL2011, Martin'11, Wang'2011, Liu'11, TopologicalQuantumBus, Martin'2012}. 
We consider here the regime where the dot is occupied by a single electon, such that, in the absence of the Majorana coupling, the system flows to the celebrated Kondo fixed point~\cite{Kondo'64}, which
has been of paramount importance to condensed matter physics~\cite{kouwenhoven2001revival}. It appears in many mesoscopic systems where an effective impurity spin is coupled to a wide range of contact materials~\cite{GlazmanJETPL'88, LeePRL'88, Kastner'98,Kondo'98, park2002coulomb, Pasupathy'04, Choi'04, Herrero'05, Hauptmann'08, Karrasch'08, Deacon'10, KretininPRB'11,LeijnsePRB'11, Golub'11, Lopez'2013, Kondo_Aguado, Pillet'13, Altland'13, BeriPRL2012, BeriPRL2013}.
In this setup, where Majorana and Kondo interactions compete at low energies, it is thus a natural and fundamental question to ask what the resulting physics is.
Furthermore, previous work
in Refs.~\onlinecite{Kondo_Aguado, Pillet'13} has shown that the competition between the superconducting proximity effect and Kondo correlations leads to the emergence of zero-bias conduction peaks at certain values of the magnetic field. Therefore, in interpreting the results of experiments on TSC-QD-NL systems, it is important to understand and distinguish the origin of zero-bias peaks.

Transport properties of TSC-QD-NL nanostructures have been investigated theoretically in Refs.~\onlinecite{LeijnsePRB'11, Golub'11, Lopez'2013}. The authors of Ref.~\onlinecite{LeijnsePRB'11} investigated transport properties of a TSC-QD-NL junction in the high-temperature limit using perturbative (in the normal-metal coupling) master equations. However, in order to understand the low-temperature properties of the TSC-QD-NL system, one needs to take into account dot-lead tunneling non-perturbatively. This aspect of the problem has been investigated by  Golub {\it et al.}~\onlinecite{Golub'11}, who concluded that Kondo correlations have a decisive effect on the transport properties. They obtain a strong temperature dependence of the zero-bias conductance which is different from a normal metal-MBS system~\cite{Fidkowski2012, Affleck'13, lutchyn_andreev13}. On the other hand, a related numerical work~\cite{Lopez'2013} considering a quantum dot coupled to two normal leads and to one end of a TSC concluded that the MBS significantly modifies the low-energy transport properties of the system in the limit of small Majorana hybridization energy. The fact that two previous publications~\cite{Golub'11, Lopez'2013} reached opposite conclusions, indicates that low energy properties of TSC-QD-NL nanostructure are not yet fully understood.

In this paper, we revisit this problem and investigate the fate of the Kondo fixed point in a TSC-QD-NL nanostructure using both analytical and numerical techniques. We show that in a wide range of physical parameters, the Majorana-induced coupling is the leading relevant perturbation and drives the system to a new infrared fixed point, i.e. the Kondo fixed point becomes unstable in the presence of Majorana-induced couplings. Thus, our conclusions regarding this issue are the opposite from those of Ref.~\onlinecite{Golub'11}, as discussed in more detail below. We also show that transport properties of the QD-based junctions involving topological and non-topological superconductors are very different. Given that quantum dots constitute relatively simple model systems with high tunability, we suggest using them as a diagnostic tool for detecting the presence or absence of localized Majorana-zero modes.

\begin{figure}[tbp]
	\centering
{\includegraphics[width=3in]{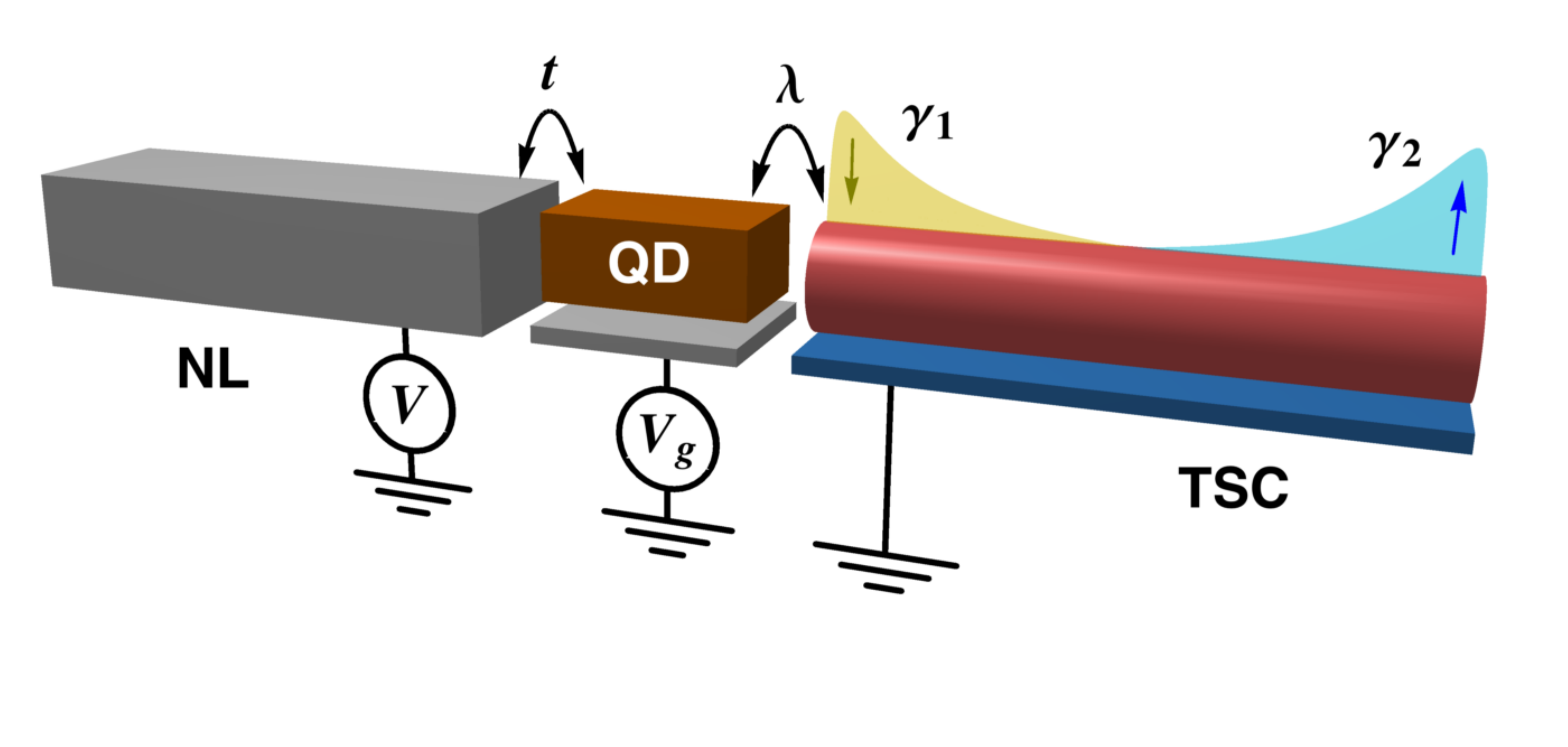}}
{\includegraphics[width=3in]{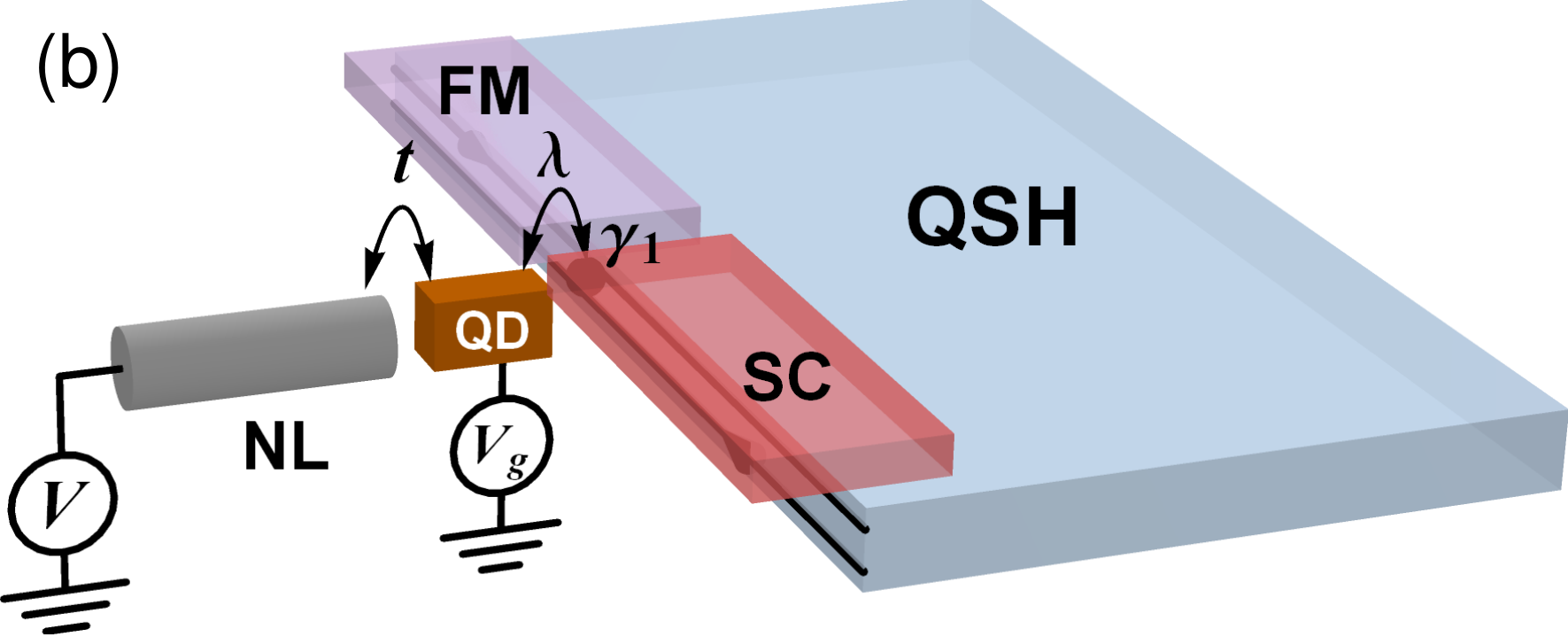}}
\caption{Schematic picture of the device: a quantum dot coupled to a localized Majorana zero mode and a normal lead. The localized Majorana mode can be realized using (a) a 1D topological superconductor or (b) a FM-SC domain wall on the edge of a quantum spin Hall insulator.  The electrochemical potential of the dot can be controlled with the gate voltage $V_g$.  We assume that the superconductor is grounded, and one can probe the low energy properties of the system through tunneling transport measurements. Here $V$ is a source-drain voltage. 
}
\label{Fig:fig1}
\end{figure}

\section{Theoretical model}

We consider a nanostructure consisting of a QD with a single spin-degenerate level coupled to a Majorana mode in a topological superconductor and an SU(2)-invariant normal lead (TSC-QD-NL junction). Our setup is sketched in Fig.~\ref{Fig:fig1}. Since we are interested in the low-energy theory valid at the energies $E \ll \Delta$, with $\Delta$ being the induced superconducting gap, the topological superconductor can be effectively described by the two Majorana zero-energy modes $\gamma_1$ and $\gamma_2$ localized at its ends. Therefore, the effective low-energy Hamiltonian for TSC-QD-NL junction reads
\begin{align}
H=&~\sum_{\sigma} \varepsilon d^\dag_{\sigma} d_{\sigma} \!+\! U n_{\uparrow}n_{\downarrow} \!+\! V \!+\! H_{\rm NL}, \label{eq:H1} \\
V=&~i  \lambda  \gamma_1 (d_{\uparrow}\!+\!d_{\uparrow}^\dag)\!+\!\sum_{\sigma}|t| \left(d_{\sigma}^\dag \psi_{\sigma}(0) \!+\!\psi_{\sigma}^\dag(0) d_{\sigma}\right), \\
H_{\text{NL}}=&- t_0 \sum_{x,\sigma} \left( \psi^\dag_\sigma(x) \psi_\sigma(x+1) + \text{h.c.} \right) \nonumber \\ &+\!U_b \sum_x n_\ua(x) n_\da(x),\label{eq:H_NL}
\end{align}
where $d_\sigma$ and $d_{\sigma}^\dag$ are annihilation and creation operators on the dot, and $n_{\sigma}=d^\dag_{\sigma}d_{\sigma}$. Here $\varepsilon$ is the chemical potential of the QD, $U$ is the strength of the electron-electron interaction on the QD, $t$ ($\lambda$) is the tunneling coupling between the lead (TSC) and the QD. We assume that the TSC is much longer than the coherence length $\xi$ and, therefore, neglect for now the coupling to other Majorana modes (i.e. $\gamma_2$ in Fig. \ref{Fig:fig1}(a)), which is exponentially small in $L/\xi$. The effect of a finite ground state degeneracy splitting in the TSC will be considered in Sec.~\ref{sec:MFSBA}. Our model defined in Eq.~\eqref{eq:H1} describes the competition between Kondo and Majorana couplings. Indeed, when the coupling to the TSC $\lambda$ is zero, the system flows to the Kondo fixed point. Turning on finite Majorana coupling $\lambda$ breaks, in addition to $U(1)$ symmetry, also time-reversal symmetry in the dot and, thus, competes with Kondo correlations.

In recent experiments~\cite{Mourik2012}, the semiconductor nanowire was made of InSb, which has a very large $g$-factor of $g_{\rm InSb}\sim 50$. The topological superconducting phase in this setup is predicted to appear at $B > B_c \approx 100$~mT, which corresponds to a Zeeman energy on the order of a Kelvin. Thus, even if there is any accidental formation of a QD in the experiment~\cite{Mourik2012}, such a large magnetic field would suppress Kondo physics in the QD. However, the situation is less clear in InAs nanowires~\cite{Das2012, Fink2012} where $g_{\rm InAs}\sim 10$ is smaller, and the Zeeman splitting at $B\sim B_c$ might be comparable with the Kondo scale~\cite{Kondo_Aguado}. In this case, it becomes non-trivial to distinguish the origin of zero-bias features, as Kondo and Majorana physics compete. To study this scenario, we consider a setup
where the semiconductor nanowire has a much larger g-factor than the normal lead and the QD, so that the magnetic field necessary to induce topological superconductivity has very little effect on the lead and the QD.  Alternatively, one can consider a localized Majorana zero-energy mode being realized at a ferromagnetic/superconducting domain wall on the edge of a quantum spin Hall (QSH) insulator, see Fig.~\ref{Fig:fig1}(b). In this case, the time-reversal symmetry is explicitly broken by a local exchange field induced by the proximity to a ferromagnetic insulator. The effect of the magnetic field generated by the ferromagnetic insulator on the QD and the lead is negligible, and one can assume that the Hamiltonian for the QD and the lead is $\mathbb{SU}(2)$-symmetric. The model Hamiltonian describing the QSH setup shown in Fig.~\ref{Fig:fig1}b is identical to the one defined in Eq.~\eqref{eq:H1}.


The Hamiltonian $H_{\text{NL}}$, defined on a lattice with hopping $t_0$, models a semi-infinite ($x\geq 0$) single-channel lead. Here $\psi_{\sigma}^\dag(x)$ and $\psi_{\sigma}(x)$ are fermion creation and annihilation operators with spin $\sigma$. In the case of a nanowire-based realizations of this setup~\cite{Mourik2012, Das2012, Deng2012, Fink2012, Churchill2013}, it might be important to take into account electron-electron interactions $U_b$ in the lead. In the continuum limit, the normal-lead Hamiltonian~\eqref{eq:H_NL} corresponds to a spinful Luttinger Liquid:
\begin{align}
H_{\rm NL}&=\sum_{j=\sigma, \rho} \frac{v_j}{2\pi} \int_0^\infty dx   \left[K_j (\nabla \theta_j)^2+ K_j^{-1} (\nabla \phi_j)^2 \right], \nonumber
\end{align}
where $v_{\rho/\sigma}$ and $K_{\rho, \sigma}$ are velocity and Luttinger liquid parameter for charge and spin modes, respectively. We follow a bosonization convention where $\psi_{r, \sigma}= \Gamma_{\sigma} e^{-\frac{i}{\sqrt 2}( r \phi_\rho-\theta_\rho)+\sigma ( r \phi_\sigma-\theta_\sigma)}/\sqrt{2\pi a}$ with $r=\pm 1$ and $\sigma=\pm 1$ for right/left-moving fermions with $\uparrow/\downarrow$ spin~\cite{Giamarchi:2003}. Here $\Gamma_{\sigma}$ is a Klein factor and $a$ is the ultraviolet cutoff of the theory. We are interested in the limit when $\varepsilon<0, U+\varepsilon>0$ favoring single occupation on the dot, and $|\lambda|$, $|t|$ both small compared to the excitation gap in the dot $\min(|\varepsilon|, U-|\varepsilon|)$. Thus, for a non-interacting lead ($K_\rho=K_{\sigma}=1$) and $\lambda=0$, the Hamiltonian~\eqref{eq:H1} corresponds to a canonical single-channel Kondo problem.

In the limit of a large charging energy on the dot, one can simplify Eq.~\eqref{eq:H1} by projecting out states with zero and double occupancy on the dot. This can be done using a Schrieffer-Wolff transformation~\cite{wolff}, see Appendix~\ref{app:der} for details. The effective Hamiltonian becomes $H=H_{\rm NL}+H_b$ with the boundary Hamiltonian $H_b$ being
\begin{widetext}
\begin{align}\label{eq:KondoMajorana}
  H_b&\!=\!-|\lambda|^2 \xi_- S_z \!+\! i\lambda |t| \gamma_1 \left \{ \frac {\xi_-}{2} (\psi_{\ua}(0)\!+\!\psi^\dag_{\ua}(0))\!+\!\xi_+ \left[(\psi_{\ua}(0)+\psi^\dag_{\ua}(0))S_z+\psi_{\da}^\dag(0) S^+ + \psi_{\da}(0) S^- \right]\right\}+|t|^2 \xi_+  \vec{s}(0) \cdot \vec{S}.
\end{align}
\end{widetext}
Here $\vec{S}$ and $\vec{s}(x)=\psi^\dag_\alpha(x){\sigma}_{\alpha\beta} \psi_\beta(x)/2$ are the impurity spin and electron spin operator at $x$; $\xi_{\pm}=\frac{1}{|\varepsilon_0|}\pm \frac{1}{U-|\varepsilon_0|}$. Different terms in Eq.~\eqref{eq:KondoMajorana} have very clear physical interpretation: the Zeeman term is generated by virtual hopping between the TSC and QD. Since the Majorana is only coupled to a spin-up electron on the dot, such a process lowers the energy of $\ua$-electrons. The second term describes tunneling of electrons between the TSC and the NL through a virtual state of the QD. The third term is the familiar Kondo interaction. One can notice that when $\varepsilon=-U/2$ (i.e. $\xi_-=0$), the boundary Hamiltonian has an additional symmetry -- particle-hole symmetry. We first analyze the generic situation $\xi_-\neq 0$ and then discuss this special case.

In the limit $\lambda\rightarrow 0$, where the system flows to the Kondo fixed point
characterized by the formation of a spin-singlet state between the localized spin on the dot and a spin of the Fermi sea,
the boundary conditions for fermions in the leads are modified. In the strong coupling limit, one finds that $\psi_R(0)=e^{2i \delta}\psi_L(0)$ with $\delta$ being a scattering phase shift, equal to $\delta=\pi/2$ in the unitary limit~\cite{Affleck1995}. On the other hand, the Majorana coupling $\lambda$ favors charge fluctuations by forming an entangled state with the fermion parity in the lead. It has been recently shown that, in the absence of a quantum dot, such a coupling drives the system to a perfect Andreev reflection fixed point characterized by the different boundary condition $\psi_R(0)=\psi^\dag_L(0)$~\cite{Fidkowski2012} for electrons in the lead. Clearly, Kondo and Majorana couplings compete with each other and drive the system to a different infrared (IR)  boundary fixed points.

\section{Results and Discussions}

\subsection{RG analysis}

In order to identify the IR fixed point the systems flows to, we study the RG flow of the boundary couplings. The minimal system of RG equations involves four couplings $h(0)=-|\lambda|^2 \xi_- $, $J_1(0)=\lambda |t|\xi_-$, $J_2(0)=\lambda |t|\xi_+$ and $J_3(0)=|t|^2\xi_+$, see Eq.~\eqref{eq:KondoMajorana}. In the weak-coupling limit $t \rightarrow 0$, we impose open boundary conditions for lead electrons, and then identify a leading relevant operator that drives the system away from the unstable fixed point. Henceforth, we assume that the normal lead has spin-SU(2) symmetry, i.e. $K_\sigma=1$. Then, the RG equations up to quadratic order in couplings are given by
\begin{equation}
 \begin{split}
  &\frac{d h}{d l }=h-\frac{J_1J_2}{4\pi v_\sigma}(1+K_{\rho}^{-1}), \\
  &\frac{d J_1}{d l }=\left(\frac{3}{4}-\frac{1}{4K_\rho}\right) J_1, \\
  &\frac{d J_2}{d l }=\left(\frac{3}{4}-\frac{1}{4K_\rho}\right) J_2-\frac{J_3 J_2}{2\pi v_\sigma}, \\
  &\frac{d J_3}{d l }=\frac{J_3^2}{2\pi v_\sigma}.
\end{split}
  \label{eq:RGflow}
\end{equation}
with $l$ being the logarithmic length scale. One can see that $h$ is relevant and tends to polarize the spin on the dot. The couplings $J_1$ and $J_2$ are relevant when $K_\rho > 1/3$, whereas the Kondo coupling $J_3$ is marginal. The competition between Majorana and Kondo interactions is reflected in the RG flow of $J_2(l)$, see second order correction proportional to $J_3 J_2$. However, in the weak coupling limit, we have $J_3/\pi v_\sigma \ll 1$ and the Majorana coupling dominates. Thus, we conjecture that the strong coupling fixed point is governed by the Majorana rather than the Kondo interaction. At the length scales $l^*$ where the Zeeman coupling becomes dominant (i.e. $h(l^*) \gg J_2(l^*)$), the spin on the QD is completely polarized along the $\hat z$-axis, and, thus, the IR fixed point corresponds to Andreev boundary condition (ABC) for spin-up electrons and normal boundary condition (NBC) for spin-down electrons, a situation which we denote as $A\otimes N$ fixed point (i.e. $\psi_{R\uparrow}^\dag(0)=\psi_{L\uparrow}(0)$ and $\psi_{R\downarrow}(0)=\psi_{L\downarrow}(0)$). The tunneling conductance through such a system is quantized in units of $2e^2/h$, similar to a TSC/Luttinger liquid (LL) junctions~\cite{Fidkowski2012, Affleck'13}. One can also understand the temperature and voltage corrections to the tunneling conductance using the previous results for TSC/LL junctions~\cite{lutchyn_andreev13}. From the structure of the lead-Majorana couplings in Eq.~\eqref{eq:KondoMajorana}, one can immediately notice that the broadening of the zero-bias conductance peak gets renormalized by the charging energy $U$. 
The decrease of the resonance width can be understood as a competition between the charging energy $U$ suppressing charge fluctuations on the dot and the coupling to the Majorana mode $\gamma_1$ favoring charge fluctuations.

The above analysis relies on perturbative RG equations which are, strictly speaking, not valid at strong coupling. Therefore, in order to access the strong coupling fixed point, we perform numerical simulations for our model. If the predictions based on the aforementioned weak-coupling analysis hold in the strong coupling limit, the system would flow to Andreev and normal boundary conditions for spin-up and spin-down, respectively. The difference in the boundary conditions for spin-up and spin-down electrons should be visible in various static correlation functions such as, for example, superconducting triplet correlation function defined as
\begin{align}\label{eq:triplet}
T_{\sigma}(x)&=\langle \psi_{\sigma}^\dag(x)\partial_x \psi_{\sigma}^\dag(x) \psi_{\sigma}(x')\partial_{x'} \psi_{\sigma} (x')\rangle|_{x'\rightarrow 0}\\
 &\propto \langle e^{2i [\theta(x)-\theta(x')]} \rangle_{x'\rightarrow 0}. \nonumber
\end{align}
One can show that at the $A \otimes N$ fixed point, $T_{\sigma}(x)$ decays as $T_{\sigma}(x)\propto |x|^{d_{\sigma}}$ where $d_{\ua}=-1/2(K_{\rho}^{-1}+1)$ and $d_{\da}=-3/2(K_{\rho}^{-1}+1)$.
Within a bosonization perspective, this can be understood as follows: Andreev boundary conditions for spin-up electrons lead to the suppression of $\theta_{\ua}$ fluctuations at the boundary, hence the decay of spin-triplet correlations is slower than in the bulk where $T_{\sigma}(x,x')\propto|x-x'|^{-(K_{\rho}^{-1}+1)}$~\cite{Giamarchi:2003}. On the contrary, $\theta_{\da}$ fluctuates strongly at the boundary and, therefore, the correlation function $T_{\da}(x)$ decays much faster than in the bulk. This drastic difference in the spin-up/down correlation function should be contrasted with the one at $\lambda=0$, where the system flows to the Kondo fixed point with $d_{\uparrow}=d_{\downarrow}$.

\begin{figure}
\centering
\includegraphics{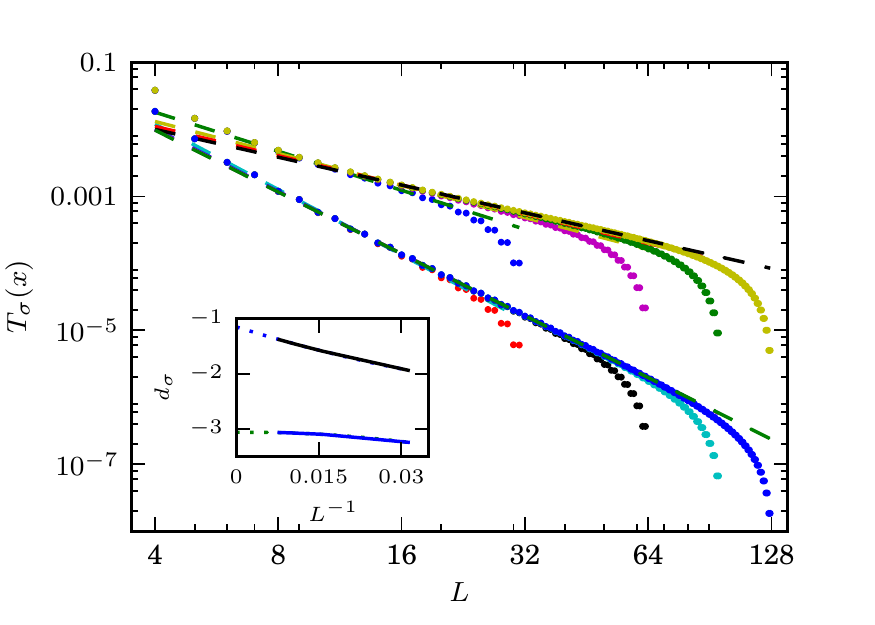}
\caption{Triplet pairing correlation function $T_\sigma(x)$ for the couplings $h = 0.2/t$, $J_1/t = 0.2$, $J_2/t = 1$, $J_3/t = 0.5$ where $t$ is the hopping in the lead. The upper set of lines show the correlation function for spin-$\ua$ fermions, while the lower set of lines show the data for spin-$\da$ fermions. Dashed lines indicate fits to a power-law decay. The inset shows the exponents $d_\ua$ (black) and $d_\da$ (blue) extracted from these fits. \label{Fig:ph_nonsymm} }
\end{figure}

To corroborate this picture, we perform simulations using the density-matrix renormalization group method~\cite{white1992,white1992-1,schollwoeck2005,schollwoeck2011}. This method has previously been applied with enormous success to many one-dimensional models, including models for the Kondo effect~\cite{SorensenPRB1996, Nishimoto2004, SorensenPRL2005, HandPRL2006, Weichselbaum2009, HolznerPRB2009}.
We perform simulations directly for the Hamiltonian~\eqref{eq:KondoMajorana}, with a real-space discretization with constant hopping for the normal lead, see Eq.~\eqref{eq:H_NL}. Such a setup allows easy access to real-space correlation functions. DMRG simulations can be systematically refined by increasing a parameter of the simulation, the so-called matrix size $M$. We perform simulations with matrix sizes up to $M=800$, ensuring accurate results also for the gapless lead. We use system sizes with an odd number of sites in the lead to allow for the ground state of the entire system to be an SU(2) singlet. Our system sizes are up to 127 sites in the lead. For the purpose of this paper, we fix the lead to half filling and set $U_b = 0$ to avoid any CDW or pairing instabilities in the lead.

Numerical results for the superconducting triplet correlation function, defined in Eq.~\eqref{eq:triplet}, are shown in Fig.~\ref{Fig:ph_nonsymm}. As one can see from the figure, the correlation function $T_{\sigma}(x)$ for spin-up and spin-down electrons decays with different exponents $d_{\uparrow}$ and $d_{\downarrow}$. From the scaling analysis  with the system size $L$, shown in the inset of the figure, we extract $d_{\uparrow}\approx - 1$ and $d_{\downarrow}\approx - 3$, which is in excellent agreement with the predictions for $A\otimes N$ boundary conditions, see the discussion above. Thus, we confirm that the strong coupling fixed point is indeed controlled by the Majorana interaction rather than the Kondo interaction, which is one of the main results of the paper.

We now emphasize the difference in transport properties for nanostructures involving topological and non-topological superconductors. In the absence of the Majorana coupling, one needs to take into account Andreev scattering at the junction,
$H\propto J_{AB}\psi_\sigma(0)\psi_{-\sigma}(0)+h.c.$.
Indeed, in the free-fermion limit ($K=1$) both Andreev and Kondo boundary couplings are marginal and compete with each other. Thus, low energy transport properties of the junction involving a non-topological superconductor (NTSC) depend on the microscopic details, i.e. the ratio of $J_3/J_{AB}$, see also discussion in Sec.\ref{sec:exp}.

\subsection{Exact solution at the particle-hole symmetric point}\label{sec:exact}
In this section we focus on the particle-hole symmetric point corresponding to a vanishing Zeeman term, i.e. $\xi_-=0$. The resulting boundary Hamiltonian is given by
\begin{align}\label{eq:KondoMajorana1}
  H_b&= i J_2  \sum_{a=x,y,z} \gamma_1 S_a \eta_a(0) + J_3 ~\vec{S} \cdot \vec{s(0)} ,
\end{align}
where $\eta_a(0)$ are Majorana operators at $x=0$ defined as $\eta_x=\psi_{\da}(0)+\psi^\dag_{\da}(0)$, $\eta_y=i(\psi^\dag_{\da}(0)-\psi_{\da}(0))$ and $\eta_z=\psi_{\ua}(0)+\psi^\dag_{\ua}(0)$. One can see that $J_2$ flows to strong coupling, and, as a result, the system forms an entangled state involving fermion parity shared between the $\gamma_1$ and $\eta_a(0)$ modes and the impurity spin. However, the nature of the boundary conditions for fermions at $x=0$ is quite non-trivial since the spin on the dot is strongly fluctuating. This can be understood from an emergent symmetry of the system. Indeed, at this special point the effective Hamiltonian is invariant under the following anti-unitary symmetry:
\begin{equation}\label{eq:symmetry}
	\tilde{\mathcal{T}}=\mathcal{C}K.
\end{equation}
Here $K$ is the complex conjugation and $\mathcal{C}$ is the charge conjugation, under which $d_\sigma \rightarrow d_\sigma^\dag$
and $\psi_\sigma \rightarrow \psi_\sigma^\dag$. One can show that $[\tilde{\mathcal{T}},H_b+H_\text{NL}]=0$. When acting on the QD impurity spin, $\tilde{\mathcal{T}}$ is similar to the time-reversal symmetry $\tilde{\mathcal{T}}\vec{S}\tilde{\mathcal{T}}^{-1}=-\vec{S}$ which implies that $\langle \vec{S}\rangle=0$. Away from the particle-hole symmetric point, both terms proportional to $\xi_-$ explicitly break  $\tilde{\mathcal{T}}$-symmetry, and induce a polarization of the impurity spin along $z$-axis.

Some insight regarding the IR fixed point in this case can be obtained when leads are non-interacting and $J_3=0$. By introducing Majorana fermion operators $\Gamma_a=2\gamma_1 S_a$, the problem can be mapped to a fermion bilinear Hamiltonian which admits an exact solution. One can show that the operators $\Gamma_a$ satisfy canonical commutation relations and anticommute with all other fermion operators, i.e. $\left\{\Gamma_a, c_\sigma \right\}=0$. We now consider the Hilbert space these matrices act upon. The Hilbert space of the original problem is given by a tensor product of the topological superconductor and the quantum dot, which is described by $4\times 4$ matrices forming a Clifford algebra. On the other hand, after the mapping we also have a Clifford algebra $\{\gamma_2, \Gamma_x, \Gamma_y, \Gamma_z \}$. The uniqueness of the Clifford algebra up to a unitary transformation ensures that the matrix representations of $\gamma_1 S_a$ and $\Gamma_a$ are equivalent. Using this mapping, one can set up a standard transport calculation~\cite{CooperPairSplitting}. We first compute the unitary scattering matrix $S(E)$ defined as
\begin{align}\label{eq:scattering}
	S(E)=1+2\pi i \hat{W}^\dag (-E-\pi i \hat{W}\hat{W}^\dag)^{-1}\hat{W},
\end{align}
where the matrix $W$ describes the coupling of Majorana modes $(\Gamma_x,\Gamma_y,\Gamma_z$) to the lead degrees of freedom:
\begin{align}
	\hat{W} = \left(\begin{array}{cccc}
      0 & iJ_x & 0 & iJ_x \\
      0 & J_y & 0 & -J_y \\
      i J_z& 0& iJ_z& 0
    \end{array}\right).
\end{align}
Here, the propagating electron and hole modes in the normal lead are described in the basis
$(\psi_\uparrow,\psi_\downarrow, \psi^\dag_\uparrow, \psi^\dag_\downarrow)$. Using the particle-hole components of the scattering matrix $P_{he}(E)$, one finds that the dc current through the system is given by
\begin{align}\label{eq:current}
I(V)&=\frac{2e}{h}\int \! dE \! \left[f(E-eV)-f(E)\right]A(E)\\
A(E)&=\sum_i|P_{he}P_{he}^\dag|_{ii}
\end{align}
where $A(E)$ is the probability of Andreev reflection, $f(E)$ is the Fermi function and $V$ is an applied bias voltage. The tunneling conductance through the junction $G=dI/dV$ at zero temperature reads
\begin{equation}\label{eq:conductancePHS}
\frac{G(V)}{2e^2/h}=\frac{W_z^2}{(eV)^2+W_z^2}+\frac{(W_x-W_y)^2 (eV)^2}{[(eV)^2+W_x^2][(eV)^2+W_y^2]}.
\end{equation}
Here the broadening width due to the coupling to Majorana modes $\Gamma_i$ is $W_i=\pi J_i^2 \nu_F$ with $i=x,y,z$ and $\nu_F$ being density of states in the lead. Since $J_x=J_y=J_z=J_2$ (see Eq.\eqref{eq:KondoMajorana1}), the second term in the above equation vanishes, and we find that the low-bias conductance is equal to $G(0)=2e^2/h$ at zero temperature, similar to the NM-MBS case~\cite{Fidkowski2012, Affleck'13}. The quantization of the conductance in units of $e^2/h$ corresponds to Andreev boundary conditions for spin-up electrons and normal boundary conditions for the spin-down electrons. Indeed, the spin-down electrons are coupled to two Majoranas, $\Gamma_x$ and $\Gamma_y$, which effectively annihilate each other, i.e. the contribution to the conductance from $\Gamma_x$ and $\Gamma_y$ is zero. Using Eqs.~\eqref{eq:current} and \eqref{eq:conductancePHS}, one finds that the temperature and voltage dependence of the conductance $G(V,T)$ at the particle-hole symmetric point is similar to that of NM-MBS junction~\cite{lutchyn_andreev13} with the width being determined by $W$. We note here that our results on the conductance at the particle-hole symmetric point are different from those of Ref.~\onlinecite{Golub'11}, who find that the zero-bias tunneling conductance has a temperature dependence which is distinct from that of the simpler NM-MBS tunnel junction~\footnote{We notice that Ref.~\onlinecite{Golub'11} also derived RG equations for the same microscopic Hamiltonian at the particle-hole symmetric point for the non-interacting lead ($K_\rho=K_\sigma=1$), but missed the tree-level scaling term for the flow of $J_2$. The RG equations in Ref.~\onlinecite{Golub'11} were obtained using field-theoretic method involving the scale invariance of the conductance. The disagreement with our momentum-shell Wilsonian RG approach originates from the fact that the coupling $J_2$ has a non-zero engineering (naive) dimension and should be rescaled to be dimensionless before computing the RG flow using the Callan-Symanzik equation. This discrepancy, however, leads to an incorrect identification of the strong coupling fixed point and invalidates the conclusions of Ref.~\onlinecite{Golub'11}.}.

\begin{figure}[tbp]
\centerline{\includegraphics{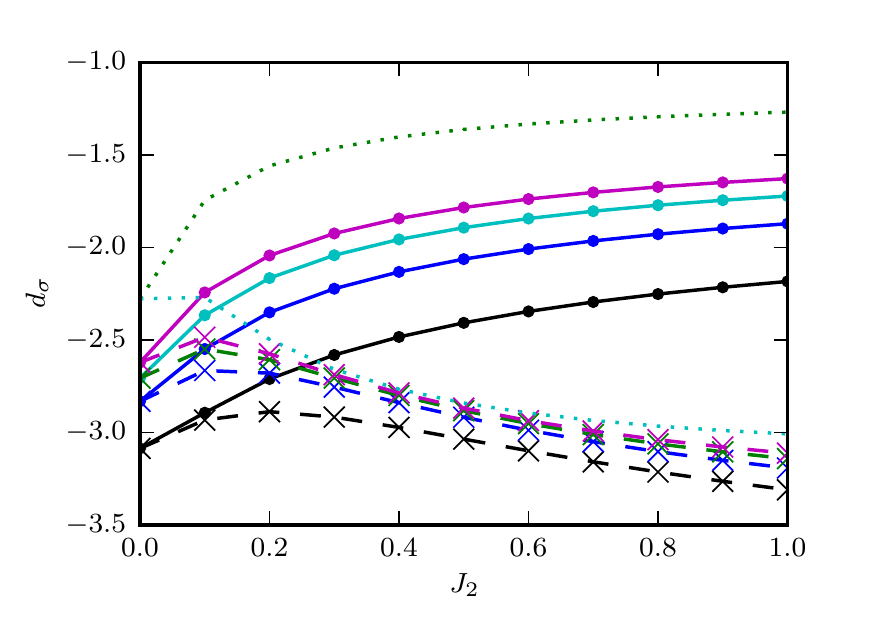}}
\caption{Dependence of the exponent $d_{\sigma}$ on the Majorana coupling $J_2$ for a fixed Kondo coupling $J_3/t = 0.5$, and $J_1 = h = 0$. Solid lines correspond to $d_\ua$, dashed lines to $d_\da$. System sizes are $L=32,64,96,128$, and an extrapolation to $L \rightarrow \infty$ is shown as dotted line. }
\label{Fig:fig3}
\end{figure}

Further insight about the strong coupling fixed point can be obtained by studying the dynamics of the impurity spin. Consider, for example, the dynamical spin-spin correlation function $\langle S_z(t)S_z(0)\rangle$. The impurity spin operator can be written in terms of Majorana operators $\Gamma_a$: $S_a=-2i\varepsilon_{abc}S_b S_c=-\frac{i}{2}\varepsilon_{abc}\Gamma_b\Gamma_c$. Then, the correlation function $\langle S_z(t)S_z(0)\rangle$ can be written as
\begin{align}
\langle S_z(t)S_z(0)\rangle &\equiv -\frac{1}{4} \langle \Gamma_x(t)\Gamma_y(t)\Gamma_x(0)\Gamma_y(0)\rangle
\end{align}
The correlation function $G(t)=\langle \Gamma_a(t)\Gamma_a(0) \rangle$ can be easily obtained by taking the Fourier transform of
\begin{equation}
	G_{a}(\omega)=\left(\omega+ \pi i W W^\dag\right)^{-1}_{aa}=\frac{1}{\omega + i W_a},
	\label{}
\end{equation}
where $a=x,y,z$. Since $J_x=J_y=J_z$, we define $W\equiv W_a$. In the long-time limit, spin-spin correlation function reads
\begin{align}
\langle S_z(t)S_z(0)\rangle|_{t\rightarrow \infty} \approx \frac{1}{4W^2 t^2}.
\end{align}
Clearly, the impurity spin operator acquires a non-trivial scaling dimension equal to one at the strong coupling fixed point, and, thus, Kondo coupling $J_3$  becomes an irrelevant perturbation in the RG sense.
We confirm our results using DMRG calculations and show that the system flows to $A\otimes N$ boundary conditions when $J_3\neq 0$, see Fig.~\ref{Fig:fig3}. One can see that for $\lambda=0$, the decay of triplet correlation functions is described by the same exponent for spin-up and spin-down electrons $d_{\uparrow}=d_{\downarrow}$. However, as soon as $\lambda\neq 0$, the exponents become different and eventually saturate at $d_{\uparrow}\approx -1$ and $d_{\downarrow}\approx -3$ for non-interacting leads. These numerical results corroborate our conjecture that in the presence of particle-hole symmetry the strong coupling fixed point is described by a new fixed point dictated by $J_2$ coupling. This fixed point corresponds to a situation where the spin on the dot is strongly entangled with the combined fermion parity of the topological superconductor and the normal lead.

Another interesting feature of the above fixed point is the dependence of the polarization of the impurity spin on the position of the energy level $\varepsilon$ on the dot. At the particle-hole symmetric point $\varepsilon = \varepsilon_0=-U/2$, the impurity spin is strongly fluctuating, i.e. $\langle S_z \rangle = 0$. If the gate voltage is detuned by $V_g$, such that $\varepsilon = \varepsilon_0 + V_g$, the spin shows a non-trivial behavior that allows us to distinguish between Majorana and Kondo physics. As the Kondo coupling does not break the $\tilde{\mathcal{T}}$-symmetry, the impurity spin has an expectation value $\langle \vec{S}\rangle =0$ for all values of $V_g$ in the absence of a coupling to the Majorana mode, $\lambda = 0$. If, on the other hand, the coupling to the Majorana mode is present, $\lambda \neq 0$,
the spin polarizes along the $\hat z$-axis for $V_g \neq 0$, see discussion after Eq.~\eqref{eq:symmetry}.
The perturbation Hamiltonian proportional to the detuning away from the particle-hole symmetric point $V_g$ is given by
\begin{align}
H_V=eV_g\left(-\frac{8|\lambda|^2}{U^2}S_z+\frac{4i \lambda |t|}{U^2}\gamma_1 \eta_z(0)\right).
\end{align}
Using linear response theory, one can now compute the spin susceptibility $\partial \langle S_z \rangle/\partial V_g \big|_{V_g = 0}$. Consider the imaginary-time dynamical spin response function:
	\begin{align}
		\chi(\tau)&=-\frac{8e\lambda^2}{U^2}\langle T_\tau S_z(\tau)S_z(0)\rangle\\
&+\frac{4ie\lambda|t|}{U^2}\langle T_\tau S_z(\tau)\gamma_1(0)\eta_z(0,0)\rangle\nonumber
	\end{align}
Using the relation $S_z=-\frac{i}{2}\Gamma_x\Gamma_y, \gamma_1=-i\Gamma_x\Gamma_y\Gamma_z$, one obtains
\begin{equation}
	\chi(\tau)=\frac{8e}{U^2}(\lambda^2-i\lambda|t|\langle\Gamma_z\eta_{z}(0)\rangle)G_{xy}(\tau)G_{xy}(-\tau),
	\label{}
\end{equation}
where the Green's function $G_{xy}(\tau)=-\langle T_\tau f(\tau)f^\dag(0)\rangle$ with $f=\frac{1}{2}(\Gamma_x+i\Gamma_y)$. Using the equations of motion, one can find $G_{xy}(i\omega)=(i\omega+iW\sgn \omega)^{-1}$.
 After some manipulations, we obtain the static spin susceptibility $\partial \langle S_z \rangle / \partial V_g \big|_{V_g = 0} \equiv \chi(\omega \rightarrow 0)$,
\begin{equation}
	\frac{\partial \langle S_z\rangle}{\partial V_g}\Big|_{V_g=0}=\frac{e}{2\pi^2\nu_F t^2}\Big(1-\frac{4\nu_F t^2}{U}\ln\frac{\Lambda}{W}\Big),
	\label{eq:susc}
\end{equation}
with
$\Lambda$ being the UV cutoff corresponding to the bandwidth in the lead and $\xi_+=\frac{4}{U}$. Close to the particle-hole symmetric point, the susceptibility $\partial\langle S_z \rangle/\partial V_g$ has a non-trivial dependence on the tunneling rates $\lambda$ and $t$.  This dependence is very distinct from the Kondo case suggesting that studies of impurity spin fluctuations in TSC-QD-NL structures might be used to help identifying Majorana zero-energy modes.

\begin{figure}
  \includegraphics{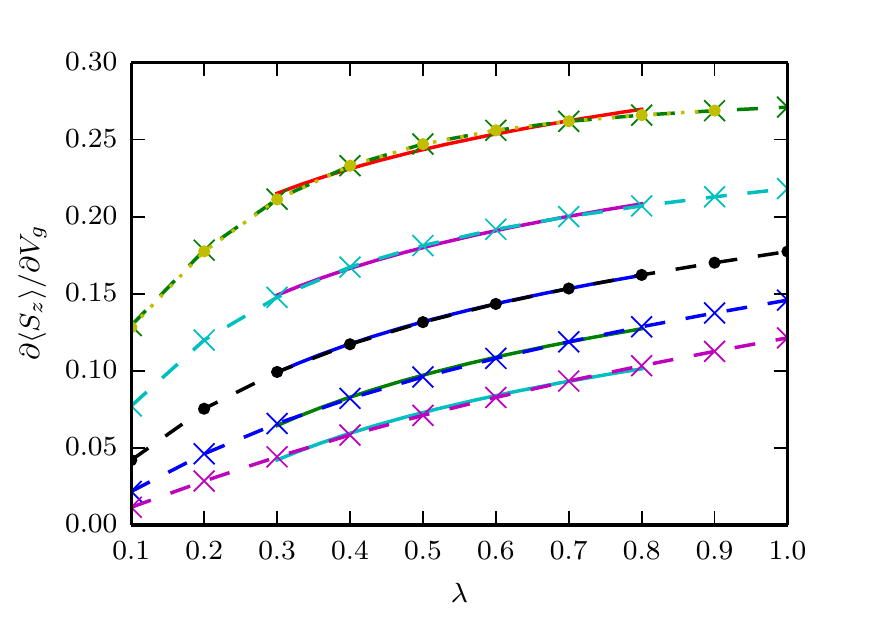}
  \caption{Impurity spin susceptibility for a system of $L=64$ sites with $U=8$ and $t=1$, where different dashed lines correspond to simulations where we have multiplied the Kondo term of Eqn.~\eqnref{eq:KondoMajorana} by $K=0,0.5,1,1.5,2$ (from top to bottom). The blue line (round points) thus represents the unmodified Hamiltonian~\eqnref{eq:KondoMajorana}. The solid lines show a fit to $a \log(b \lambda^2)$ (cf. Eqn.~\eqnref{eq:susc}), with $a$, $b$ fit parameters, over a regime of intermediate $\lambda$. The dotted, yellow line is a comparison to results obtained in linear response for the case without Kondo coupling. \label{fig:susc} }
\end{figure}

To numerically confirm this behavior, we resort to DMRG simulations of the system. We simulate systems
up to $L=64$ sites described by Hamiltonian~\eqnref{eq:KondoMajorana}, where we set
$U=8$ and $t=1$. For our simulations, we use $M=400$ states and we measure $\langle S_z \rangle$
on the impurity. We slightly detune the chemical potential on the quantum dot from the particle-hole symmetric
point and perform simulations in the range $V_g \in [-0.05,0.05]$, allowing us to numerically
extract the derivative $\partial \langle S_z \rangle / \partial V_g$. To gain additional insight into the crossover
between Kondo- and Majorana-dominated physics, we modify the Hamiltonian~\eqnref{eq:KondoMajorana}
by multiplying the Kondo coupling term by an additional factor $K$ to either suppress or enhance the Kondo
energy scale.
Our results are shown in Fig.~\ref{fig:susc}. We find that for intermediate values of $\lambda$ and in particular the physical limit $K=1$, the susceptibility can be well fit by a $\log \lambda^2$, as suggested by \eqref{eq:susc}. If $\lambda\sim 1$ or $\lambda$ is much smaller than the level spacing in the lead, finite-size effects become significant and our previous calculation in the thermodynamic limit is no longer applicable. We have checked that a modified theory on finite-size lattice systems reproduces the numerical data very well.
We observe that enhancing the Kondo coupling suppresses the spin susceptibility. This is easily understood by considering that in the Kondo-dominated strong-coupling fixed point, i.e. for $\lambda \ll t$, the impurity spin forms a singlet with lead electrons, i.e. $\langle S \rangle \rightarrow 0$ and, thus, the susceptibility vanishes.

\subsection{Results away from the particle-hole symmetric point: slave-boson mean field theory}\label{sec:MFSBA}

The results discussed in the previous section are valid close to the particle-hole symmetric point. We now develop a theory away from this point in the large-$U$ limit. In this case one can use a slave boson approximation~\cite{Coleman84, Newns87} proven to successfully capture the strong correlation effect in the Anderson impurity models, see, e.g., Ref.~\onlinecite{Nagaosa_book}. In order to take into account the interplay between the Kondo correlation and the Majorana coupling, we go back to the initial Anderson-type model of the TSC-QD-NL junction and project out the doubly occupied state using a mean-field slave boson approximation (MFSBA). A similar study was carried out in Ref.~\onlinecite{Golub'11} but, as explained below, the non-trivial mean-field solution corresponding to the Majorana-dominated regime, which is the main result of our paper, was not treated there. To make this section self-contained, we briefly review the slave boson approach and discuss mean field solutions in different parameter regime. Technical details are presented in the Appendix~\ref{app_slave}.

We begin by writing the Anderson Hamiltonian for the TSC-QD-NL junction:
\begin{align}
H=&~i\delta\gamma_1\gamma_2+\sum_{\sigma} \varepsilon d^\dag_{\sigma} d_{\sigma} \!+\! U n_{\uparrow}n_{\downarrow} \!+\! V \!+\! H_{\rm NL}, \\
V=&~i  \lambda  \gamma_1 (d_{\uparrow}\!+\!d_{\uparrow}^\dag)\!+\!\sum_{k\sigma}|t| \left(d_{\sigma}^\dag \psi_{k\sigma} \!+\!\text{h.c.}\right), \\
H_{\text{NL}}=&\sum_{k\sigma}\xi_k\psi_{k\sigma}^\dag\psi_{k\sigma}.
\end{align}
where $\psi_{k\sigma}$ are fermion annihilation operators in the normal lead, and $\delta$ is the ground state degeneracy splitting in the TSC due to finite size effects. In the infinite-$U$ limit, double occupancy of the QD is suppressed. To this end we introduce new operators $d_{\sigma}\rightarrow f_{\sigma}b^\dag$ and $d^\dag_{\sigma}\rightarrow f^\dag_{\sigma}b$ where  $b$ and $f_{\sigma}$ represent unoccupied and singly-occupied states, respectively. Double occupancy is excluded by introducing the constraint $b^\dag b+\sum_{\sigma}f_{\sigma}^\dag f_{\sigma}=1$. Thus, the effective action of the system reads
\begin{widetext}
\begin{align}\label{eq:Sslave}
S_{\rm sb}=&\int d\tau \left[\sum_\sigma f_\sigma^\dag (\partial_\tau+\varepsilon) f_\sigma+\sum_{k\sigma}\psi_{k\sigma}^\dag (\partial_\tau+\xi_k)  \psi_{k\sigma} + i  \lambda  \gamma_1 (f_{\uparrow}b^\dag\!+\!f_{\uparrow}^\dag b)\!+\!\sum_{k\sigma}|t| \left(f_{\sigma}^\dag  \psi_{k\sigma} b \!+\!\text{h.c.}\right)+\sum_{i=1,2}\gamma_i \partial_\tau \gamma_i \right. + i \delta \gamma_1 \gamma_2\nonumber\\
&+\left. \eta(b^\dag b+\sum_\sigma f^\dag_\sigma f_\sigma-1)\right],
\end{align}
\end{widetext}
where the last term enforces the constraint on the Hilbert space. We now make a mean-field approximation and replace boson operators by their expectation value $\langle b \rangle =\langle b^\dag \rangle=b$, which together with the Lagrange multiplier $\eta$ are going to be determined self-consistently by minimizing the action $S_{\rm sb}$:
\begin{align}\label{eq:self}
\frac{\partial S_{\rm sb} }{\partial \eta}\!&=0 \rightarrow	b^2+\sum_\sigma\langle f^\dag_\sigma f_\sigma\rangle=1\\
\frac{\partial S_{\rm sb} }{\partial b}\!&\!=\!0\! \rightarrow	2b\eta\!+\!t\sum_{k\sigma}\!\left(\langle f^\dag_\sigma \psi_{k\sigma}\rangle\!+\!\text{c.c}\right)\!+\!i\lambda \left\langle \gamma\left(f_\uparrow^\dag\!+\!f_\uparrow\right)\right\rangle\!=\!0\nonumber
\end{align}
The calculation of the above correlation functions is presented in the Appendix~\ref{app_slave}. Here we highlight our main results.

\begin{figure}[b]
	\centering
	\includegraphics[width=0.9\columnwidth]{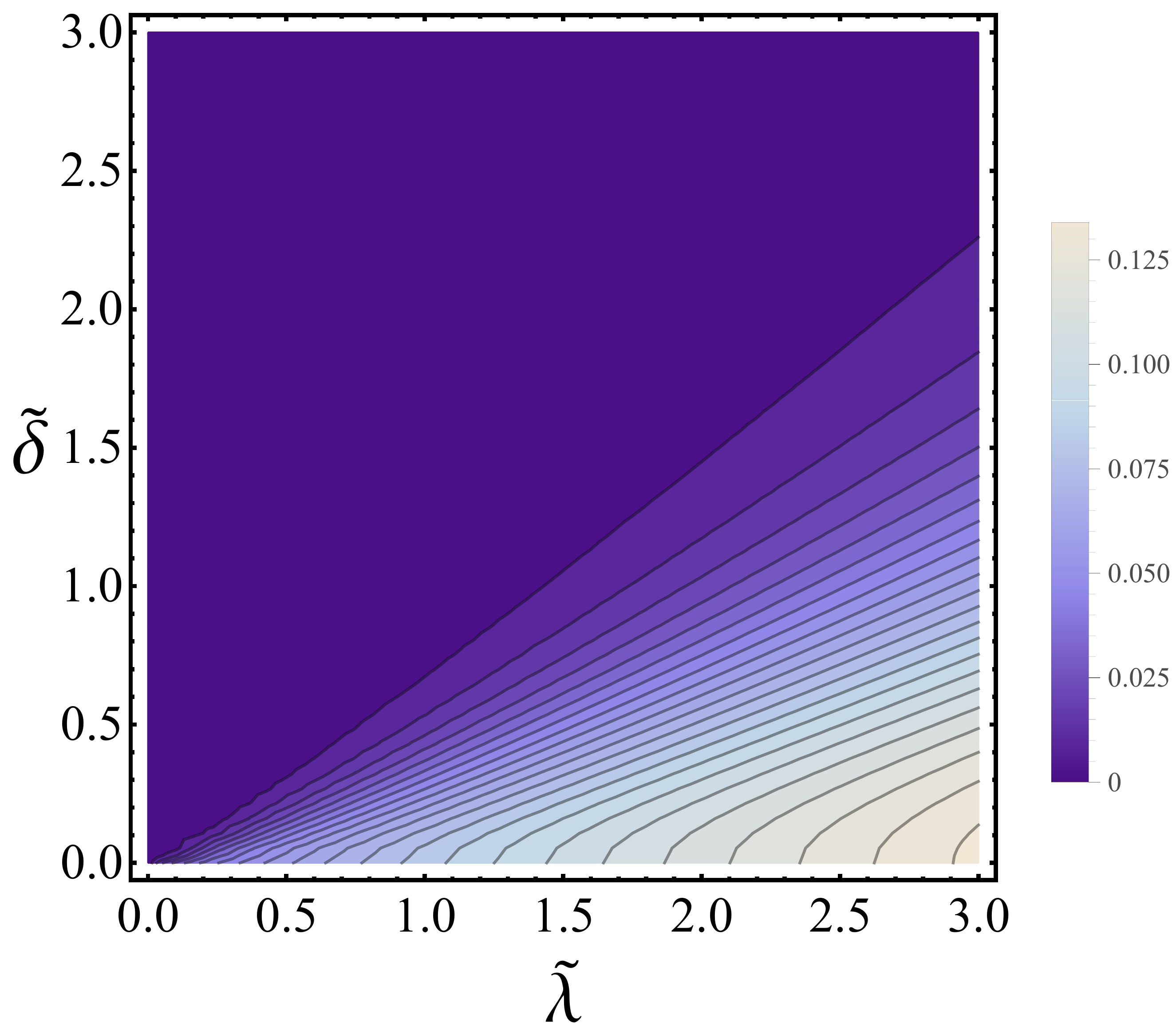}
	\caption{The mean-field solution for $b$ as a function of $\tilde{\lambda}$ and $\tilde{\delta}$ obtained by numerically using Eqs.~\eqref{eq:self}. We have set $\varepsilon=-6\Gamma$ and the bandwidth $\Lambda=30\Gamma$ in the calculation.}
	\label{fig:bsol}
\end{figure}

The general solution can be obtained numerically as well as, in some cases, analytically. We first consider the case of no splitting $\delta=0$. In the single-occupancy limit where $\varepsilon<0$ and $|\varepsilon| \gg \lambda, |t|$, the probability of a QD being empty $b^2$ is small. Thus, one can solve the self-consistency equations assuming $b\rightarrow 0$. From the first equation above, we find that $|\varepsilon+\eta|\sim \Gamma b^4$ with $\Gamma=\pi |t|^2 \nu_F$ being the lead-induced broadening in the dot. Using the expression for $\eta$ and evaluating the integrals in the second equation, one finds
\begin{equation}\label{eq:selfcons2}
	\eta+\frac{2\Gamma}{\pi}\ln\frac{\Gamma b^2}{\Lambda}-\frac{\lambda}{2\sqrt{2}b}=0.
\end{equation}
Here we kept only leading terms in the expansion in small $b$; $\Lambda$ is a UV cutoff corresponding to the bandwidth in the lead. When $\lambda=0$, we recover the solution for the Kondo model
\begin{align}
	b^2\approx\frac{\Lambda}{\Gamma}\exp\left(-\frac{\pi|\varepsilon|}{2\Gamma}\right),
	\label{eqn:kondo_b}
\end{align}
and $\Gamma b^2=\Lambda e^{-\frac{\pi |\varepsilon|}{2\Gamma}}\equiv T_K$ corresponds to the Kondo temperature $T_K$. When $\lambda$ is large, the second term in Eq.~\eqref{eq:selfcons2} is more important since it is more divergent. Thus, an approximate solution reads
\begin{equation}
	b\approx\frac{\lambda}{2\sqrt{2}|\varepsilon|}.
	\label{}
\end{equation}
This is a new mean-field solution of the self-consistency equations which was not considered in Ref.~\onlinecite{Golub'11}. Once again, we see that the Majorana coupling is a more relevant perturbation than the Kondo coupling and, thus, determines the low energy properties of the system. The crossover between the two regimes can be determined by matching the two solutions for $b$:
\begin{equation}
	\lambda_c \sim \sqrt{\frac{T_K}{\Gamma}} |\varepsilon|=\sqrt{\frac{\Lambda}{\Gamma}} |\varepsilon| \exp\left(-\frac{\pi |\varepsilon|}{4\Gamma}\right).
	\label{}
\end{equation}
Thus, the Majorana-dominated regime corresponds to $\lambda \gg \lambda_c $ with the value of $b$ determined by the Majorana coupling whereas $\lambda \ll \lambda_c $ corresponds to the Kondo-dominated regime.

We will now discuss finite size effects in the nanowire by assuming that Majorana degeneracy splitting $\delta$ is non-zero. In this case, we find solutions of  Eqs.~\eqref{eq:self} numerically. A plot of $b$ as a function of the rescaled parameters $\tilde{\delta}=\delta/\Gamma, \tilde{\lambda}=\lambda/\Gamma$ is shown in Fig.~\ref{fig:bsol}. One can see that when $\delta$ is large, the value of $b$ closely tracks Eq.~\eqref{eqn:kondo_b}. As $\delta$ is increased there is a crossover between the Majorana-dominated regime and the Kondo-dominated regime. As expected, the splitting energy $\delta$ reduces the parameter space corresponding to the Majorana-dominated regime.


\subsection{Quantum-dot tunneling experiments}\label{sec:exp}

Having solved the mean-field equations for $b$ and $\eta$, we can now compute experimentally observable quantities. In this section, we study the differential tunneling conductance, having in mind a setup similar to that of Ref.~\onlinecite{Mourik2012}, see Fig.~\ref{Fig:fig1}. At the qualitative level one can already see that, once the parameters $\eta$ and $b$ are determined within the MFSBA, the dynamics of spin-up and -down electrons decouple. Spin-down electrons do not contribute to Andreev reflection since they are not coupled to the Majorana mode $\gamma_1$.\footnote{Here we neglect Andreev scattering terms proportional to $d_{\uparrow}d_{\downarrow}+h.c.$ which are suppressed by the large superconducting gap.} Thus, the mean-field results are consistent with the boundary conditions $A_{\uparrow} \otimes N_{\downarrow}$. Using the scattering matrix formalism outlined in Sec.\ref{sec:exact}, see Eq.\eqref{eq:scattering}, one can find the probability of Andreev reflection:
\begin{widetext}
\begin{align}\label{eq:PH_prob}
	A(E)\!=\!\frac{b^8 \Gamma^2 E^2 \lambda ^4}{b^4 E^2 \lambda ^4 \left(\Gamma^2 b^4+E^2\right)\!-\!2 b^2 E^2 \lambda ^2 (E^2 \!-\! 4 \delta^2 )\left(\Gamma^2 b^4\!+\!E^2\!-\!\tilde \varepsilon^2\right)\!+\!(E^2\!-\!4 \delta^2)^2 \left(\left(\Gamma^2 b^4\!+\!\tilde \varepsilon^2\right)^2\!+\!E^4\!+\!2E^2 (\Gamma^2 b^4 -\tilde \varepsilon^2)\right)}
\end{align}
\end{widetext}
where $\tilde \varepsilon=\varepsilon+\eta$. The differential tunneling conductance can be obtained with the help of Eq.~\eqref{eq:current}, and is given by $G(V)=\frac{2e^2}{h}A(eV)$ at zero temperature. One can see that the functional dependence $G(V)$ is different from the Lorentzian form characteristic to the simpler TPSC-NL structures and has a much richer structure. At zero splitting $\delta=0$, the conductance is still quantized at zero bias $G(0)=2e^2/h$ but the broadening of the resonance is a non-trivial function of various parameters:
\begin{align}
G(V)=\frac{2e^2}{h}\frac{b^8 \Gamma^2 \lambda ^4}{ [(eV)^2+\Gamma^2 b^4][((eV)^2-\lambda^2 b^2)^2+(eV)^2\Gamma^2 b^4]}
\end{align}
The value of $b$ in Majorana- and Kondo-dominated regimes is $b\approx \lambda/\sqrt 8 |\varepsilon|$ and $b \approx \sqrt{T_K/\Gamma}$, respectively. At small bias $V \rightarrow 0$, one can estimate the effective width of the resonance $\Gamma_{\rm eff}$ to be
\begin{equation}
\Gamma_{\rm eff}\approx
	\begin{cases}
 \frac{\Gamma \lambda^2}{8|\varepsilon|^2} & \mbox{ for }  \lambda \gg \lambda_c\\
{\rm min} \left \{T_K, \frac{\lambda^2}{\Gamma} \right \}  & \mbox{ for }  \lambda \ll \lambda_c,
\end{cases}.
	\label{}
\end{equation}

\begin{figure}[!htb]
	 \centering
	 \includegraphics[width=\columnwidth]{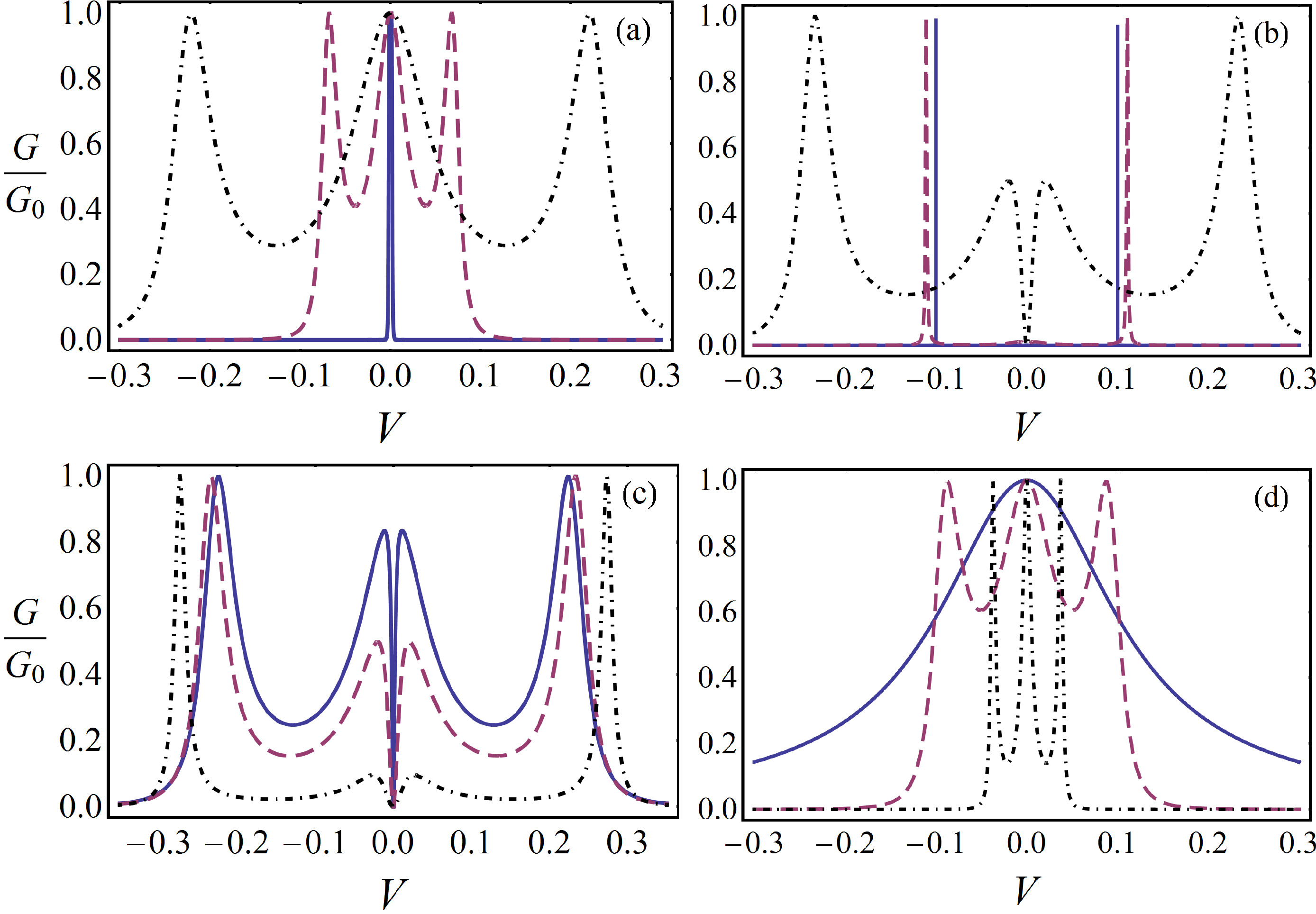}
	 \caption{Zero-temperature tunneling conductance $G(V)$ as a function of various parameters. Here all energies  are re-scaled by $\Gamma$, $G_0=2e^2/h$. (a) $\delta=0, \varepsilon=-3.5$, $\lambda=0.05$(solid), $\lambda=0.5$(dashed) and $\lambda=1.0$(dashed-dot) (b) the same as (a) but $\delta=0.1$. (c) $ \lambda=1, \varepsilon=-3.5$, $\delta=0.05$(solid), $\delta=0.1$(dashed) and $\delta=0.2$(dashed-dot) (d) $ \lambda=0.5, \delta=0.1$, $\varepsilon=-1$(solid), $\varepsilon=-3$(dashed) and $\varepsilon=-5$(dashed-dot).}
	 \label{fig:cond}
 \end{figure}

These results can be understood as follows. In the Majorana-dominated regime, the physics of the QD is determined by the Majorana strong coupling fixed point. Therefore, the width of the resonance is proportional to $t^2 \lambda^2 \nu_F/|\varepsilon|^2 \sim W$. This is consistent with the results at the particle-hole symmetric point, see Sec.~\ref{sec:exact}. The Kondo-dominated regime corresponds to the small-$\lambda$ limit, where the Majorana mode $\gamma_1$ is localized in the TSC and is only weakly coupled to the QD. The effective width of the zero-bias peak is determined by the smaller of the two rates $T_K$ and $\lambda^2/\Gamma$, and therefore is much sharper than in the Majorana-dominated regime where the width of the resonance is suppressed only as a power law in $|\varepsilon|$. Another interesting feature in the tunneling conductance is the appearance of the sidebands as shown in Fig.~\ref{fig:cond}(a). These sidebands originate from the splitting of the Kondo resonance by the induced Zeeman term, Eq.~\eqref{eq:KondoMajorana}.
In our model coupling to the TSC breaks the $U(1)$ charge conservation as well as the time-reversal symmetry. Both these effects lead to the suppression of the Kondo effect. In the Majorana-dominated regime finite-bias resonances appear at $eV=\pm \lambda b\sim \lambda^2/|\varepsilon|$~\footnote{Note that we have neglected the effect of an external magnetic field $B$ here assuming that the $g$-factor in the QD is small. If Zeeman splitting due to an external $B$ is larger than the exchange-induced one then the position of the sidebands will be determined by the Zeeman term. We believe this is the case in the recent Majorana experiments~\cite{Mourik2012}.}; the width of these resonances is of the order of $W$.

We note that the aforementioned dependence on $\lambda$ is very different from the one in the s-wave SC-QD-NL junction~\cite{Deacon'10, Kondo_Aguado, Pillet'13} where Majorana interaction is absent ($\lambda=0$).
This can be understood most easily in the limit of a large superconducting gap $\Delta \rightarrow \infty$ (i.e. $\Delta \gg \Gamma_S, \, \Gamma, \, T_K$), where one can integrate out SC degrees of freedom. In the low energy approximation, the effect of an s-wave superconductor can be represented as Andreev scattering term $H_P=\Gamma_S d_{\uparrow}^\dag d_{\downarrow}^\dag +h.c.$. This term breaks $U(1)$ symmetry and competes with the Kondo singlet state. It has been shown that the Abrikosov-Suhl resonance in the tunneling conductance, present at small $\Gamma_S\ll \Gamma$, gets gradually suppressed with the increase of the coupling to a superconductor, see Refs.~\cite{Clerk'00, Domanski'08, review_Kondo_SC} for details. This behaviour should be contrasted with $G(V)$ shown in Fig.\ref{fig:cond} where the zero-bias peak becomes more pronounced with increasing of the coupling to TSC. We note, however, that in the non-perturbative regime (i.e. $\Delta \sim \Gamma_S, \, \Gamma, \, T_K $) the situation is more complicated due to the possible appearance of the Shiba-like bound states induced by an unpaired electron spin in the QD~\cite{Zitko_arxiv}. If Shiba levels are close to the Fermi energy, one would expect to observe an enhancement in the subgap tunneling conductance. Thus, it is important to control the coupling and keep it small, $\Gamma_S \ll \Delta$, so that the contribution of the Shiba states to the zero-bias tunneling conductance is suppressed. Overall, we find that there is a wide parameter regime $T_K \ll \Gamma_S/\lambda \ll \Delta$ where the dependence of the zero-bias peak on the coupling to the superconductor is very different for topological and non-topological states allowing one to distinguish between Kondo and Majorana physics.

We now discuss effect of the splitting energy $\delta \neq 0$ due to a finite size of the TSC. Analytical results are not particularly illuminating in this case, and we present numerical solution instead. The plots of the tunneling conductance as a function of various parameters are shown in Fig.\ref{fig:cond}b-d. Overall, many qualitative features can be understood as a convolution of the local density of states in the TSC and QD. At small $\lambda$ the splitting energy leads to the emergence of the two sharp peaks at energies $\pm \delta$, see Fig.\ref{fig:cond}b, the width of these peaks becomes larger with the increase of $\lambda$. Eventually, when $\lambda^2/|\varepsilon|\gg \delta$, the shape of the conductance $G(V)$ changes qualitatively: two sideband peaks located at $eV\sim\lambda^2/|\varepsilon|$ emerge. In this limit, the position of the peaks and their width are weakly dependent on $\delta$, compare Fig.\ref{fig:cond}a and b. The splitting, however, strongly affects the zero-bias feature and eliminates the zero-bias peaks entirely.

\begin{figure}
	 \centering
	 \includegraphics[width=0.8\columnwidth]{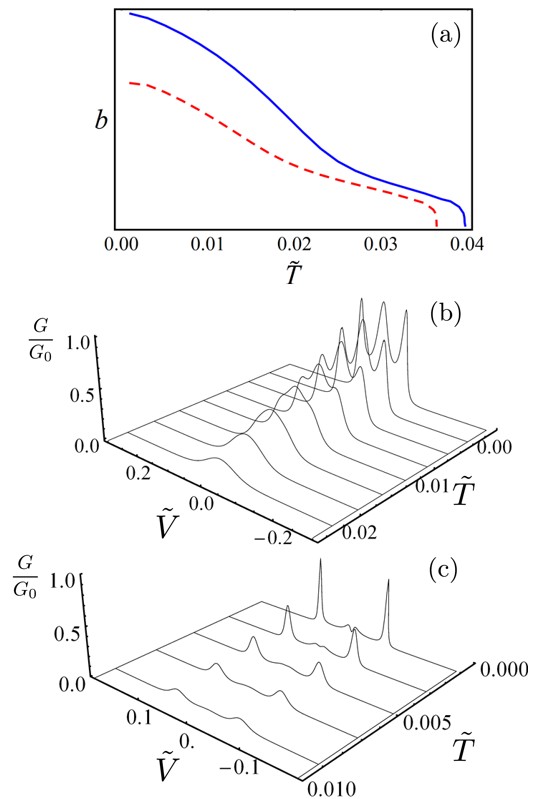}
	 \caption{Dependence of the tunneling conductance $G(V, T)$ on temperature and voltage bias. (a) The solution of the variational parameter $b$ as a function of temperature $T$. Here all the energy scales are re-scaled by $\Gamma$; the parameters are $\varepsilon=-3.5, \lambda=0.5$, and $\delta=0$ (solid) and $\delta=0.1$ (dashed). The other variational parameter $\eta$ is weakly dependent on temperature. Panel (b) and (c) are the plots of the tunneling conductance $G(V, T)$ as a function of temperature $T$ and voltage bias $V$ for $\delta=0$ and $\delta=0.1$, respectively.}
	 \label{fig:cond2}
\end{figure}

So far we have considered zero temperature limit $T=0$. Using Eq.~\eqref{eq:current} and \eqref{eq:PH_prob}, we now calculate the temperature dependence of the differential tunneling conductance. The plot of $G(V,T)$ is shown in Fig.~\ref{fig:cond2}. The triple peak structure in Fig.~\ref{fig:cond2}a gets smeared by the temperature, and one might have to go to very small temperatures in order to observe this feature, especially in the small $\lambda$-limit (i.e. Kondo-dominated regime). 
In the case of a finite splitting $\delta \neq 0$, the width of the peak becomes more narrow. As a result, tunneling conductance is suppressed even faster by thermal fluctuations, see Fig.\ref{fig:cond2}c.

\section{Conclusions}
In this paper we study physical properties of a quantum dot in the Coulomb blockade regime coupled to a 1D topological superconductor and a normal lead. In the experimentally relevant parameter regime, the low energy theory for such a system involves Kondo and Majorana-induced interactions. We study the competition between Kondo and Majorana couplings, and show that they drive the system to different many-body ground states. In the universal limit, where the splitting of the ground state degeneracy $\delta$ associated with the topological superconductor is zero, we show that the infrared fixed point is governed by a Majorana-induced coupling rather than the Kondo one. When $\delta\neq 0$, we discuss the crossover between the Kondo- and Majorana-dominated regimes as a function of various physical parameters, such as couplings between the quandum dot and the superconductor and normal lead, the splitting energy $\delta$ as well as the gate voltage determining the single-particle energy in the dot.

By considering the impurity spin susceptibility and the differential tunneling conductance in TSC-QD-NL junction, which both are experimentally accessible quantities, we show how one can distinguish between Kondo and Majorana physics in the lab. In particular, we predict that when a coupling to the Majorana mode is present, the impurity spin polarization $\langle \mathbf{S} \rangle $ exhibits a strong dependence on tuning the gate voltage away from the particle-hole symmetric point: while it vanishes at the PHS point, the spin becomes polarized as the gate voltage is detuned. This has to be contrasted with the Kondo-dominated regime where the impurity spin is disordered for all values of detuning from the particle-hole symmetric point. Quantitatively, the spin susceptibility shows a non-trivial dependence on the Majorana coupling. Furthermore, Majorana signatures in TSC-QD-NL junction should also manifest in various time-dependent experiments, see, for example, a recent proposal on how to detect universal non-equilibrium signatures of Majoranas in quench dynamic~\cite{Moore'14}.

We also discuss the zero-bias anomaly in the tunneling conductance and show how one can distinguish between the Kondo and Majorana features in realistic experimental settings. Although both Kondo and Majorana correlations might lead to zero-bias anomaly in $dI/dV$, we show that the dependence of the zero-bias peak on various parameters, such as, for example, the coupling between the QD and the superconductor, is quite different. Our results have important implications for the experiments trying to detect Majorana zero modes since the nature of the many-body ground state for a quantum dot coupled to topological and non-topological superconductors is very different. We believe that the exceptional degree of the parameter control in quantum dot experiments might prove to be very useful for disentangling different phenomena in the laboratory. In particular, the present setup where the transmission between TSC and NL can be tuned by the charging energy as well as the couplings to the lead and superconductor might be quite useful to reduce the subgap density of states, also known as the ``soft gap'' problem, which appears due to the hybridization of the states in the nanowire with the NL; for more details, see e.g. Ref.~\cite{Stanescu_soft}.

\acknowledgements
We thank C.M. Marcus, A. Mitchell and C. Nayak for stimulating discussions and Y. Avishai for helpful correspondence. We acknowledge the hospitality of the Max Planck Institute for the Physics of Complex Systems in Dresden, where part of this work has been performed. M.B. thanks Microsoft Station Q for hospitality in the initial stages of this work and acknowledges funding through the Institutional Strategy of the University of Cologne within the German Excellence Initiative. The DMRG code was developed with support from the Swiss Platform for High-Performance and High-Productivity Computing (HP2C) and based on the ALPS libraries~\cite{bauer2011-alps}.

\appendix

\section{Derivation of an effective Hamiltonian for topological superconductor-quantum dot-normal lead junction}\label{app:der}

In this section we derive an effective Hamiltonian for a topological superconductor-quantum dot-normal lead junction, see Fig.~\ref{Fig:fig1}. We assume that the quantum dot is in the Coulomb-blockade regime, and the Hamiltonian for the system is given by Eq.~\eqref{eq:H1}. The charging energy on the dot can be tuned with the gate voltage $V_g$. In order to study the Kondo effect, we will focus on the regime of a single electron occupancy on the dot, i.e. $\varepsilon_0<0, \varepsilon_0+U>0$ as measured with respect to the Fermi energy in the lead. To derive the low-energy Hamiltonian of the system, we perform a Schrieffer-Wolff transformation~\cite{wolff} and eliminate zero- and double-occupancy sectors. The projectors to the subspace of $n$ electrons on the quantum dot $P_n$ are given by
\begin{subequations} \begin{align}
  P_0&=(1-n_\uparrow)(1-n_\downarrow),\\
  P_1&=(1-n_\uparrow)n_\downarrow+(1-n_\downarrow)n_\uparrow, \\
  P_2&=n_\uparrow n_\downarrow,
  \label{}
\end{align} \end{subequations}
Then, the effective Hamiltonian for the system can be formally written as
\begin{equation}
  H_\text{eff}=H_{11}+\sum_{n=0,2}H_{1n}\frac{1}{E-H_{nn}}H_{n1}.
  \label{}
\end{equation}
We assumed here that the typical energies involved have $|E| \ll \min(|\varepsilon_0|, U-|\varepsilon_0|)$. After some algebra, one finds explicit expressions for $H_{01}$ and $H_{12}$:
\begin{subequations} \label{eq:matrix_el} \begin{align}
  H_{01}&=P_0 H P_1=\sum_\sigma (t\psi_\sigma^\dag+i\lambda_\sigma\gamma)(1-n_{-\sigma})d_{\sigma}\\
  H_{12}&=P_1 H P_2=\sum_\sigma (t\psi_\sigma^\dag+i\lambda_\sigma\gamma) n_{-\sigma}d_\sigma.
\end{align} \end{subequations}
Using Eqs.~\eqref{eq:matrix_el}, one obtains the following contributions to the effective Hamiltonian:
\begin{align}
  \! H_{12}\frac{1}{E\!-\!H_{22}}H_{21}&\!\approx\!\sum_{\sigma, \sigma'}\left[\frac{(t^2\psi_\sigma^\dag \psi_{\sigma'}\!+\!\lambda_\sigma\lambda_{\sigma'}^*)d_\sigma n_{-\sigma} d_{\sigma'}^{\dag}n_{-\sigma'}}{U\!-\!|\varepsilon_0|}\right.\\
  &+\left.\frac{it\gamma(\lambda_{\sigma'}^*\psi_\sigma^\dag+\lambda_\sigma \psi_{\sigma'}) d_\sigma n_{-\sigma} d_{\sigma'}^{\dag}n_{-\sigma'}]}{U\!-\!|\varepsilon_0|}\right] \nonumber\\
H_{10}\frac{1}{E-H_{00}}H_{01}&\approx -\frac{1}{|\varepsilon_0|} \sum_{\sigma\sigma'}[t^2 \psi_\sigma \psi_{\sigma'}^\dag+\lambda_\sigma^*\lambda_{\sigma'}\\
&-it\gamma(\lambda_\sigma^* \psi_{\sigma'}^\dag+\lambda_{\sigma'}\psi_\sigma)]d_\sigma^\dag\overline{n}_{-\sigma} d_{\sigma'}\overline{n}_{-\sigma'}]\nonumber
  \label{}
\end{align}
Here we approximate $E-H_{22}\approx |\varepsilon_0|-U$ and $E-H_{00}\approx -|\varepsilon_0|)$  in the denominator for the virtual intermediate states assuming that the couplings $|t|, |\lambda_\sigma|$ are small $|t|, |\lambda_\sigma|\ll \min(|\varepsilon_0|, U-|\varepsilon_0|)$. Given that the lead Hamiltonian is $\mathbb{SU}(2)$-spin invariant, one can, without loss of generality, simplify the above expressions by choosing a quantization axis for the coupling between the Majorana and the QD. By setting $\lambda_\downarrow=0$ and $\lambda_{\uparrow}=\lambda$, the effective boundary Hamiltonian becomes
\begin{align}\label{eq:eff_Hamilt}
  H_{b}&=\xi_-\left[\frac{it\lambda}{2} \gamma(\psi_\uparrow+\psi_\uparrow^\dag)-|\lambda|^2 S^z\right]+\xi_+t^2 \vec{s}(0)\cdot\vec{S}\\
  &+\xi_+t\lambda\left[i\gamma(\psi_\uparrow+\psi_\uparrow^\dag)S^z+i\gamma(\psi_\downarrow S^+ + \psi_\downarrow^\dag S^-)\right].\nonumber
\end{align}
where $\vec{S}$ and $\vec{s}(0)=\psi^\dag_\alpha(0){\sigma}_{\alpha\beta} \psi_\beta(0)/2$ are the impurity spin and electron spin operator at $x=0$. The coefficients $\xi_\pm$ are defined as
\begin{equation}
   \xi_\pm =\frac{1}{|\varepsilon_0|}\pm\frac{1}{U-|\varepsilon_0|}.
  \label{}
\end{equation}
The effective Hamiltonian~\eqref{eq:eff_Hamilt} is the main result of this section. The physical meaning of different terms in Eq.~\eqref{eq:eff_Hamilt} has been explained in the main text.

\section{Derivation of the RG flow equations}
In this Appendix we provide details of the derivation of the RG equations. The imaginary-time partition function of the Hamiltonian defined in Eq.~\eqref{eq:KondoMajorana} can be written as a path integral:
\begin{equation}
  \mathcal{Z}=\int\mathcal{D}\theta_\rho\mathcal{D}\theta_\sigma\, e^{-(S_0+S_b)}.
  \label{}
\end{equation}
The effective action $S_0$ for the boundary field $\theta_{\sigma/\rho}(\tau)$ at $x=0$ is obtained by integrating bulk degrees of freedom
\begin{equation}
  S_0=\sum_{\lambda=\rho,\sigma}\frac{K_\lambda}{2\pi}\int\frac{\di\omega}{2\pi}|\omega||\theta_\lambda(\omega)|^2.
  \label{}
\end{equation}
The boundary action $S_b$ is given by
\begin{align}
   \mathcal{S}_b&=\int d\tau \left\{h S^z+\frac{iJ_1\gamma_1 \Gamma_\uparrow}{\sqrt{2\pi a}}  \cos\left(\frac{\theta_\rho+\theta_\sigma}{\sqrt{2}}\right)\right.\\
  &+\frac{2i\gamma_1}{\sqrt{2\pi a}}\left[\Gamma_\uparrow J_2^z S^z\cos\left(\frac{\theta_\rho+\theta_\sigma}{\sqrt{2}}\right)\right.\nonumber\\
  &+\left. \frac{J_2^\perp}{2}\Gamma_\downarrow \left(S^+ e^{\frac{i}{\sqrt{2}}(\theta_\sigma-\theta_\rho)}+\text{h.c.}\right)\right]\nonumber\\
  &\left.+\frac{J_3^zS^z}{2\sqrt{2}\pi v_\sigma}i\partial_\tau\theta_{\sigma}+\frac{J_3^\perp}{4\pi a}(S^+\Gamma_\uparrow\Gamma_\downarrow e^{i\sqrt{2}\theta_\sigma}+\text{h.c.})\right\}\nonumber
  \label{}
\end{align}
Here $\Gamma_\sigma$ are Klein factors and $a$ is the ultraviolet cutoff length scale. We introduced different notations for $J_3^z$, $J_3^\perp$, $J_2^z$ and $J_2^\perp$ for convenience of deriving RG equations. However, because of symmetry in the microscopic model, we set $J_2^z=J_2^\perp$ and $J_3^z=J_3^\perp$ at the end of the calculation. One should keep in mind that $K_\sigma=1$ for the $\mathbb{SU}(2)$-invariant normal lead.

We now perform weak-coupling RG analysis in the frequency domain. We first separate the fields $\theta_\lambda$ into fast and slow modes: $\theta_\lambda=\theta_\lambda^<+\theta_\lambda^>$ with $\theta_\lambda^<$ and $\theta^>$ containing the modes with frequencies $0<|\omega|<\frac{\Lambda}{b}$ and $\frac{\Lambda}{b}<|\omega|<\Lambda$, respectively. After integrating over the fast modes, the new effective action can be calculated using cumulant expansion:
\begin{equation}
  S_\text{eff}[\theta^<]=S_0[\theta^<]+\langle S_b\rangle-\frac{1}{2}(\langle S_b^2\rangle-\langle S_b\rangle^2).
  \label{}
\end{equation}
Here $\langle\dots\rangle$ denotes integrating out the fast modes. The first-order term $\langle S_b\rangle$ in the expansion gives familiar renormalization of the couplings at the tree level. We skip the details for brevity since the results can be easily obtained and concentrate on second-order corrections to $J_i$ instead. The derivation of these corrections is rather lengthy so we break this Appendix into several subsections. To simplify the notations, we define:
\begin{equation}
  \concomp{A}{B}\equiv\langle AB\rangle-\langle A\rangle\langle B\rangle.
  \label{}
\end{equation}

\subsubsection{Evaluation of the contribution from $J_3^\perp J_2^z$ term}

Let us consider the contribution originating from $J_3^\perp J_2^z$ term:
\begin{widetext}
  \begin{align}\label{eqn:rg_jzjp1}
\delta S^{(a)}&=-\frac{1}{2}\frac{{i J_3^\perp J_2^z}}{(2\pi a)^{\frac 3 2}} \!\int d\tau d\tau' S^+(\tau)\Gamma_\uparrow(\tau)\Gamma_\downarrow(\tau)\gamma_1 \Gamma_\uparrow(\tau')S^z(\tau')\concomp{e^{\sqrt{2}i\theta_\sigma(\tau)}}{\cos\frac{\theta_\rho(\tau')\!+\!\theta_\sigma(\tau')}{\sqrt{2}}}
\\
&\!=\!-\frac{1}{2}\frac{{iJ_3^\perp J_2^z}}{(2\pi a)^{\frac 3 2}} \!\int d\tau d\tau' S^+(\tau)\Gamma_\uparrow(\tau)\Gamma_\downarrow(\tau)\gamma_1 \Gamma_\uparrow(\tau')S^z(\tau')\frac{e^{\sqrt{2}i\theta_\sigma^<(\tau)}}{2}\!\nonumber\\	
\!&\!\times\!\left[\!e^{\frac{i}{\sqrt{2}}[\theta_\sigma^<(\tau')\!+\!\theta^<_\rho(\tau')]}\langle e^{\frac{i}{\sqrt{2}}\theta_\rho^>(\tau')}\rangle\concomp{e^{\sqrt{2}i\theta_\sigma^>(\tau)}}{e^{\frac{i}{\sqrt{2}}\theta_\sigma^>(\tau')}}
	\!+\!e^{-\frac{i}{\sqrt{2}}[\theta_\sigma^<(\tau')\!+\!\theta_\rho^<(\tau')]}\langle e^{-\frac{i}{\sqrt{2}}\theta_\rho^>(\tau')}\rangle\concomp{e^{\sqrt{2}i\theta_\sigma^>(\tau)}}{e^{-\frac{i}{\sqrt{2}}\theta_\sigma^>(\tau')}}
	\right]\nonumber
\end{align}
\end{widetext}
Using the following identities
\begin{align}
	\langle S^a(\tau)S^b(\tau')\rangle &=\frac{1}{4}\delta_{ab}+\frac{i}{2}\varepsilon^{abc}S^c\text{sgn}(\tau-\tau')\\
\langle \Gamma(\tau)\Gamma(\tau')\rangle&=\text{sgn}(\tau-\tau')
\end{align}
the expressions for the spin and Majorana operators can be simplified to
\begin{equation}
  S^+(\tau)\Gamma_\uparrow(\tau)\Gamma_\downarrow(\tau)\gamma \Gamma_\uparrow(\tau')S^z(\tau')=\frac{1}{2}S^+\gamma\Gamma_\downarrow
  \label{}
\end{equation}

Next, we evaluate the following correlation functions:
\begin{align}
&\langle e^{\frac{i}{\sqrt{2}}\theta_\rho^>(\tau)}\rangle=e^{-\frac{1}{4}\langle {\theta_\rho^>}(\tau)^2\rangle }=e^{-\frac{1}{4K_\rho}\ln b}=b^{-\frac{1}{4K_\rho}}
\end{align}
and
\begin{align}\label{app:corr1}
&\concomp{e^{\sqrt{2}i\theta^>_\sigma(\tau)}}{e^{\pm\frac{i}{\sqrt{2}}\theta_\sigma^>(\tau')}}=e^{-\langle{\theta_\sigma^>}(\tau)^2\rangle-\frac{1}{4}\langle{\theta_\sigma^>}(\tau')^2\rangle}\nonumber\\
&\times\Big(e^{\mp\langle\theta_\sigma^>(\tau)\theta_\sigma^>(\tau')\rangle}-1\Big)
\end{align}
Let us study the expression for $\langle\theta_\lambda^>(\tau)\theta_\lambda^>(\tau')\rangle$ where $\lambda$ can be $\rho$ or $\sigma$:
\begin{align}
  \begin{split}
  g_\lambda(\tau\!-\!\tau')\equiv&\langle\theta_\lambda^>(\tau)\theta_\lambda^>(\tau')\rangle=\frac{1}{K_\lambda}\int_{\Lambda/b}^\Lambda\frac{\di \omega}{\omega}\cos{\left[\omega(\tau\!-\!\tau')\right]}\\
  &
  \begin{cases}
  \approx\frac{1}{K_\lambda}K_0\Big(\frac{\Lambda|\tau\!-\!\tau'|}{b}\Big) & |\tau-\tau'|\gg b/\Lambda\\
  =\frac{1}{K_\lambda}\ln b & \tau=\tau'
\end{cases}
\end{split}
  \label{}
\end{align}
where $K_0(\tau)$ being the zero-th order modified Bessel function. The function $K_0(\tau)$ is peaked at $|\tau|\ll b/\Lambda$ and decays exponentially for $|\tau|\gg b/\Lambda$. Therefore, it is reasonable to limit the integral over $|\tau-\tau'|$ by an upper cutoff $b/\Lambda$. We then make a change of the variables and define
\begin{equation}
  T=\frac{\tau+\tau'}{2}, s=\tau-\tau'.
  \label{}
\end{equation}
Now Eq. \eqref{eqn:rg_jzjp1} becomes
\begin{align}
\delta S^{(a)}&=\frac{{-iJ_3^\perp J_2^z}}{8(2\pi a)^{\frac 3 2}} \int_0^\beta\di T\,\gamma_1\Gamma_\downarrow S^+e^{\sqrt{2}i\theta_\sigma^<(T)} \\
\int_{-b/\Lambda}^{b/\Lambda}&\di s\, \Big[e^{\frac{i}{\sqrt{2}}[\theta_\sigma^<(T)+\theta^<_\rho(T)]}b^{-\frac{1}{4K_\rho}-\frac{5}{4}}(e^{-g_\sigma(s)}-1)\nonumber\\ &+e^{-\frac{i}{\sqrt{2}}[\theta_\sigma^<(T)+\theta_\rho^<(T)]}b^{-\frac{1}{4K_\rho}-\frac{5}{4}}(e^{g_\sigma(s)}-1)\Big]\nonumber
\end{align}
Since $g_\sigma(s)$ is strongly peaked at $s=0$, we replace the integrand $e^{\pm g_\sigma(s)}$ with $e^{\pm g_\sigma(0)}$.

Collecting all the terms, one finds that the term in \eqref{eqn:rg_jzjp1} is evaluated to
\begin{widetext}
 \begin{equation}
 \delta S^{(a)}=\frac{-i{J_3^\perp J_2^z}}{8(2\pi a)^{\frac 3 2}} \frac{2b}{\Lambda} b^{-\frac{1}{4K_\rho}-\frac{5}{4}}\int_0^\beta\di T\gamma_1 \Gamma_\downarrow S^+ \left[e^{\frac{i}{\sqrt{2}}[\theta_\sigma^<(T)-\theta_\rho^<(T)]}(b-1)- e^{\frac{i}{\sqrt{2}}[3\theta_\sigma^<(T)+\theta_\rho^<(T)]}(b^{-1}-1)\right]
  \label{}
\end{equation}
\end{widetext}
The second term in the integrand corresponds to the generation of a new term which is irrelevant under RG, and thus can be ignored. Thus, the second-order correction reads as
 \begin{equation}
\delta S^{(a)}\!=\!f(b)\int_0^\beta\di T\,\frac{-i{J_3^\perp J_2^z \gamma_1\Gamma_\downarrow S^+}}{8\pi\sqrt{2\pi a}v_\sigma} e^{\frac{i}{\sqrt{2}}[\theta_\sigma^<(T)\!-\!\theta_\rho^<(T)]}+h.c.
  \label{eqn:rg_jzjp2}
\end{equation}
Here we used $a\Lambda=v_\sigma$; the function $f(b)$ is  defined as
\begin{equation}
  f(b)=b^{-\frac{1}{4}(1+\frac{1}{K_\rho})}(b-1).
  \label{}
\end{equation}
We note that in order to obtain RG flow of $J_2^\perp$, one has to take into account factor of 2 in Eq.~\eqref{eqn:rg_jzjp2}  coming from switching $\tau$ and $\tau'$ in \eqref{eqn:rg_jzjp1}.

One can compute the contribution of $S^- S^z$ term in a similar fashion. As expected, it is given by the Hermitian conjugate of \eqref{eqn:rg_jzjp2}. Finally, the sum of these two terms gives the second-order correction to the $J_2^\perp$ term.

\subsubsection{Evaluation of the contribution from $J_3^\perp J_2^\perp$ term}
We now compute the correction proportional to $J_3^\perp J_2^\perp$
\begin{widetext}
  \begin{align}
\delta S^{(b)}&=-\frac{i}{4}\int d\tau d\tau' \frac{J_3^\perp J_2^\perp}{(2\pi a)^{\frac 3 2}}S^+(\tau)\Gamma_\uparrow(\tau)\Gamma_\downarrow(\tau)\gamma_1\Gamma_\downarrow(\tau') S^-(\tau') e^{\sqrt{2}i\theta_\sigma^<(\tau)}e^{\frac{i}{\sqrt{2}}[\theta_\rho^<(\tau')-\theta_\sigma^<(\tau')]}\langle e^{\frac{i}{\sqrt{2}}\theta_\rho^>(\tau')}\rangle \concomp{e^{\sqrt{2}i\theta_\sigma^>(\tau)}}{e^{-\frac{i}{\sqrt{2}}\theta_\sigma^>(\tau')}}\nonumber\\
&+ S^-S^+ { \rm term}
	\label{eq:app2}
  \end{align}
\end{widetext}
The correlation function involving spin and Majorana operators evaluates to
 \begin{equation}
   \begin{gathered}
   S^+(\tau)\Gamma_\uparrow(\tau)\Gamma_\downarrow(\tau)\gamma_1\Gamma_\downarrow(\tau') S^-(\tau')\\
   =\Big(\frac{1}{2}\text{sgn}(\tau-\tau')+S^z\Big)\gamma_1\Gamma_\uparrow.
   \end{gathered}
  \label{eqn:rg_spinop}
\end{equation}
The connected correlation function is given by
\begin{equation}
  \concomp{e^{\sqrt{2}i\theta_\sigma^>(\tau)}}{e^{-\frac{i}{\sqrt{2}}\theta_\sigma^>(\tau')}}=b^{-\frac{5}{4}}( e^{g_\sigma(\tau-\tau')}-1).
  \label{}
\end{equation}
By rewriting the integral~\eqref{eq:app2} in terms of $T$ and $s$ variables and integrating over $s$, one obtains
  \begin{equation}
\delta S^{(b)}=-i\int_0^\beta\di T\, \frac{J_3^\perp J_2^\perp f(b)}{2\pi v_\sigma \sqrt{2\pi a}} S^z \gamma_1\Gamma_\uparrow \cos\left({\frac{\theta_\rho^<(T)+\theta_\sigma^<(T)}{\sqrt{2}}}\right)
	\label{}
  \end{equation}
Once again one has to take into account factor of 2 to obtain RG flow of $J_2^z$ coming from switching $\tau$ and $\tau'$ in \eqref{eq:app2}

We would like to point out that in \eqref{eqn:rg_spinop}, $\gamma\Gamma_\uparrow$ is also generated which in principle contributes to the renormalization of $J_1$ coupling. However, due to $\text{sgn}(\tau-\tau')$ the integral vanishes. Thus, there are no corrections to $J_1$-coupling at this order.

\subsubsection{Evaluation of the contribution from $J_3^z J_2^\perp $ term}
We now compute the contribution proportional to $J_3^z J_2^\perp$
\begin{align}
\delta S^{(c)}=&\frac{1}{4}\int d\tau d\tau'\frac{J_3^z J_2^\perp}{2\pi \sqrt{\pi a}v_\sigma}S^z(\tau)\gamma_1\Gamma_\downarrow(\tau')S^+(\tau')\nonumber\\
& \times \concomp{\partial_\tau \theta_\sigma}{ e^{\frac{i}{\sqrt{2}}[\theta_\sigma(\tau')-\theta_\rho(\tau')]}}.
	\label{eqn:rg_jzj2p}
\end{align}
The calculation of the connected correlation function can be done in two steps. First, we introduce fields $\theta_\lambda^>(\tau)$ and $\theta_\lambda^>(\tau)$ and finds that the relevant correlation function is given by
\begin{equation}
  \begin{split}
  \concomp{\partial_\tau \theta_\sigma}{ e^{\frac{i}{\sqrt{2}}[\theta_\sigma(\tau')-\theta_\rho(\tau')]}}&=\\
  e^{\frac{i}{\sqrt{2}}[\theta_\sigma^<(\tau')-\theta_\rho^<(\tau')]}\langle e^{-\frac{i}{\sqrt{2}}\theta^>_\rho(\tau')}\rangle
&\langle\partial_\tau \theta_\sigma^> e^{\frac{i}{\sqrt{2}}\theta_\sigma^>(\tau')}\rangle
\end{split}
  \label{}
\end{equation}
Here we used the fact that $\langle \partial_\tau\theta_\sigma^>\rangle=0$. In order to calculate the correlation function $\langle\partial_\tau \theta_\sigma^> e^{\frac{i}{\sqrt{2}}\theta_\sigma^>(\tau')}\rangle$ we use the following identity:
\begin{equation}
\partial_\tau\theta_\sigma^>(\tau)=\lim_{\epsilon\rightarrow 0}\frac{1}{i\epsilon}\partial_\tau e^{i\epsilon\theta_\sigma^>(\tau)},
  \label{}
\end{equation}
and rewrite the correlation function as
\begin{equation}
  \begin{split}
  \langle\partial_\tau \theta_\sigma^> e^{\frac{i}{\sqrt{2}}\theta_\sigma^>(\tau')}\rangle=
  \lim_{\epsilon\rightarrow 0}\frac{1}{i\epsilon}\partial_\tau e^{-\frac{1}{{4}}[\sqrt{2}\epsilon\theta_\sigma^>(\tau)+\theta_\sigma^>(\tau')]^2}\\
  =\frac{i}{\sqrt{2}}\partial_\tau \langle\theta_\sigma^>(\tau)\theta_\sigma^>(\tau')\rangle e^{-\frac{1}{4}\langle{\theta_\sigma^>}^2(\tau')\rangle}.
  \end{split}
  \label{}
\end{equation}
 Then, Eq.~\eqref{eqn:rg_jzj2p} becomes
\begin{align}
\delta S^{(c)}&=\frac{i}{b^{\frac{1}{4}+\frac{1}{4K_\rho}}} \int_0^\beta\frac{\di T  J_3^z J_2^\perp}{16\pi \sqrt{2\pi a}v_\sigma}\gamma_1\Gamma_\downarrow S^+ e^{\frac{i}{\sqrt{2}}[\theta_\sigma^<(T)-\theta_\rho^<(T)]}  \nonumber\\
   &\times   \int\di s\, \text{sgn}(s)\partial_s g_\sigma(s)\label{eq:app3}\\
   &=-i\int_0^\beta\di T\,\frac{J_3^z J_2^\perp \gamma_1\Gamma_\downarrow S^+ }{8\pi v_\sigma \sqrt{2\pi a}}\frac{\ln b}{b^{\frac{1}{4}+\frac{1}{4K_\rho}}}e^{\frac{i}{\sqrt{2}}[\theta_\sigma^<(T)-\theta_\rho^<(T)]}\nonumber.
\end{align}
The other term $S^zS^-$ yields the Hermitian conjugate of Eq.~\eqref{eq:app3}. At the end of the day, we find
 \begin{align}
\delta S^{(c)}\!=\!-\frac{\ln b}{b^{\frac{1}{4}\!+\!\frac{1}{4K_\rho}}}\int_0^\beta\frac{\di T\, J_3^z J_2^\perp}{8\pi v_\sigma \sqrt{2\pi a}}i\gamma_1\Gamma_\downarrow (S^+e^{i\frac{\theta_\sigma^<(T)\!-\!\theta_\rho^<(T)}{\sqrt{2}}}\!+\!{\rm h.c.}).
  \label{}
\end{align}
One has to multiply above expression by the factor of $2$ due to switching $\tau$ and $\tau'$.

\subsubsection{System of RG equations}
To obtain the system of RG equations we collect all the terms, and then rescale the imaginary time parameter $\tau$ to $\tau'=\tau/b$ which leads to an additional factor of $b$ in all corrections. Now one can expand $b$ in term of the small parameter $\delta \Lambda/\Lambda$, i.e.  $b\approx 1+\frac{\delta\Lambda}{\Lambda}$. The Taylor expansion of the function $f(b)$ is $f(b)\approx \frac{\delta\Lambda}{\Lambda}$. Comparing the original action with the cumulant expansion, we finally obtain the following RG equations for $J_2^\perp$ and $J_2^z$:
\begin{align}
	\frac{d J_2^\perp}{dl}&=\left(\frac{3}{4}-\frac{1}{4K_\rho}\right)J_2^\perp-\frac{J_3^\perp J_2^z}{4\pi v_\sigma}-\frac{J_2^\perp J_3^z}{4\pi v_{\sigma}}\\
\frac{d J_2^z}{dl}&=\left(\frac{3}{4}-\frac{1}{4K_\rho}\right)J_2^z-\frac{J_3^\perp J_2^\perp}{2\pi v_{\sigma}}
\end{align}
In addition to these RG equations, we need to consider flow of $J_1$ and $J_3$ couplings. As mentioned above, the coupling $J_1$ is not renormalized by the Kondo interaction. The flow of Kondo coupling is not affected by the Majorana couplings at this order of perturbative RG equations. Using similar calculations, we find second-order corrections to the RG flow of $h$, see Eq.\eqref{eq:RGflow}. Finally, combining all the terms and taking into account the symmetry of the microscopic Hamiltonian (i.e. $J_2^z=J_2^\perp$ and $J_3^z=J_3^\perp$) we arrive at Eqs.~\eqref{eq:RGflow}.

\section{Green's functions in the slave-boson mean-field theory}\label{app_slave}

The mean-field slave boson action for the system is defined in Eq.\eqref{eq:Sslave}. Within this approximation, the Hilbert space constraint enforced by $\eta$ is satisfied at the mean-field level which simplifies the calculation. We now define the following Green's functions
\begin{equation}
	\begin{gathered}
	G_1(\tau)=-\langle T_\tau\gamma_1(\tau)f_{\uparrow}(0)\rangle\\
	G_{f\sigma}(\tau)=-\langle T_\tau f_{\sigma}^\dag(\tau)f_{\sigma}(0)\rangle\\
	G_T(k\sigma,\tau)=-\langle T_\tau \psi_{k\sigma}^\dag(\tau)f(0)\rangle\\
	\end{gathered}
	\label{}
\end{equation}
and compute them using equation of motion technique to find
\begin{align}
	G_1(\omega_n)
	&=\frac{-\lambda b}{\omega_n^2+\delta^2+\frac{2\lambda^2b^2\omega_n(\omega_n+\Gamma_n)}{(\omega_n+\Gamma_n)^2+\tilde \varepsilon^2}}\frac{\omega_n}{i\omega_n-\tilde \varepsilon+i\Gamma_n}\\
	G_{f\sigma}(\omega_n)&=\frac{1+i\lambda_{\sigma} b G_1(\omega_n)}{i\omega_n-\tilde \varepsilon+i\Gamma_n}\\
	G_T(k\sigma,\omega_n)&=\frac{tb}{i\omega_n-\xi_k}\frac{1+i\lambda_\sigma b G_1(\omega_n)}{i\omega_n-\tilde \varepsilon+i\Gamma_n}
\end{align}
where $\tilde \varepsilon=\varepsilon+\eta$, $\lambda_{\uparrow}=\lambda$, $\lambda_{\downarrow}=0$ and $\Gamma_n=\Gamma b^2 \sgn \omega_n$ with $\Gamma=\pi |t|^2 \nu_F$. Thus, the self-consistency equations determining $b$ and $\eta$ are given by
\begin{align}
b^2&-\frac{1}{\beta}\sum_{\sigma, n}G_{f\sigma}(\omega_n)e^{i\omega_n 0^+}=1\label{eq:mean1}\\
2b\eta&-\frac{2t}{\beta}\sum_{k\sigma,n} {\rm Re}[ G_{T}(k\sigma,\omega_n)e^{i\omega_n 0^+}]\label{eq:mean2}\\
&-\frac{2\lambda}{\beta}\sum_{n} {\rm Re}[i G_1(\omega_n)e^{i\omega_n 0^+}]=0\nonumber
\end{align}
We now compute these correlation functions at zero temperature:
\begin{widetext}
	\begin{align}
		\sum_{\sigma}n_\sigma & =-\frac{1}{\beta}\sum_{\sigma, n}G_{f\sigma}(\omega_n)e^{i\omega_n 0^+}=-\frac{1}{\beta}\sum_{n}e^{i\omega_n 0^+}\left[\frac{2}{i\omega_n-\tilde \varepsilon+i\Gamma_n}+\frac{ i\lambda b G_1(\omega_n)}{i\omega_n-\tilde \varepsilon+i\Gamma_n}\right]\\
&=1-\frac{2}{\pi}\arctan\frac{\tilde \varepsilon}{\Gamma b^2}+8\tilde \varepsilon\lambda^2b^2\int_0^\infty\frac{\di \omega}{2\pi}\frac{\omega(\omega+\Gamma b^2)}{\left(\omega^2+\delta^2+\frac{2\lambda^2b^2\omega(\omega+\Gamma b^2)}{(\omega+\Gamma b^2)^2+{\tilde \varepsilon}^2}\right)[(\omega+\Gamma b^2)^2+\tilde \varepsilon^2]^2}
\end{align}
Taking into account above expression, one can show that the solution of Eq.\eqref{eq:mean1} is $\tilde \varepsilon \approx C_1\Gamma b^4$ with $C_1$ being a constant of order one. Thus, in the limit of small $b$ assumed here $\eta\approx -\varepsilon+O(b^4)$.
\begin{align}\label{eq:int2}
t\sum_k\langle\psi_{k\sigma}^\dag f_\sigma\rangle &=-\frac{t}{\beta}\sum_{k,n} G_{T}(k,\omega_n)e^{i\omega_n 0^+}=-\frac{|t|^2 b}{\beta}\sum_{k,n}\frac{1}{i\omega_n-\xi_k}\left[ \frac{1}{i\omega_n-\tilde \varepsilon+i\Gamma_n}+\frac{i\lambda_{\sigma} b G_1(\omega_n)}{i\omega_n-\tilde \varepsilon+i\Gamma_n}\right]
\end{align}
Let's consider the first term in Eq.\eqref{eq:int2}
\begin{align}
	\frac{|t|^2 b}{\beta}\sum_{k,n}\frac{1}{i\omega_n-\xi_k}\frac{1}{i\omega_n-\tilde \varepsilon+i\Gamma_n}&=-\frac{\Gamma b }{\beta}\sum_{n}\frac{i \sgn \omega_n}{i\omega_n-\tilde \varepsilon+i\Gamma_n}=-\frac{\Gamma b}{\pi} \int_{-\Lambda}^0 \di \omega \frac{\omega-\tilde \varepsilon}{(\omega-\tilde \varepsilon)^2+(\Gamma b^2)^2}\approx\frac{\Gamma b}{\pi}\ln\frac{\Lambda}{\Gamma b^2},
\end{align}
where we have introduced UV cutoff $\Lambda$ to regularize the integral. One can show that the second term does not have any UV divergences and thus is much smaller than the first term and can be neglected. Finally, we compute the last term in Eq.\eqref{eq:mean2}:
\begin{align}
\langle\gamma_1f_\uparrow\rangle &= -\frac{1}{\beta}\sum_n G_1(\omega_n)e^{i\omega_n 0^+}=-2i\lambda b\int_0^\infty\frac{\di \omega}{2\pi}\frac{\omega(\omega+\Gamma b^2)}{\left(\omega^2+\delta^2+\frac{2\lambda^2b^2|\omega|(|\omega|+\Gamma b^2)}{(|\omega|+\Gamma b^2)^2+\tilde \varepsilon^2}\right)[(\omega+\Gamma b^2)^2+\tilde \varepsilon^2]}\nonumber\\
			&\stackrel{\tilde \varepsilon \rightarrow 0}{\approx}-2i\lambda b\int_0^\infty\frac{\di \omega}{2\pi}\frac{\omega}{(\omega^2+\delta^2)(\omega+\Gamma b^2)+{2\lambda^2b^2\omega}}\stackrel{\delta\rightarrow 0}{=}-2i\lambda b\int_0^\infty\frac{\di \omega}{2\pi}\frac{1}{\omega(\omega+\Gamma b^2)+{2\lambda^2b^2}}\nonumber\\
&=
\begin{cases}
	-\frac{4i}{\sqrt{8\lambda^2-\Gamma^2 b^2}}\left(\frac{\pi}{2}-\arctan\frac{\Gamma |b|}{\sqrt{8\lambda^2-\Gamma^2 b^2}}\right) & 8\lambda^2>\Gamma |b|\\
	- \frac{2i\lambda}{\sqrt{\Gamma^2b^2-8\lambda^2}} \ln\frac{\Gamma |b|+\sqrt{\Gamma^2b^2-8\lambda^2}}{\Gamma |b|-\sqrt{\Gamma^2b^2-8\lambda^2}} & 8\lambda^2<\Gamma |b|
\end{cases}.
	\end{align}
\end{widetext}

%

\end{document}